\documentclass[11pt,a4]{article}
\pdfoutput=1
\usepackage{jheppub}
\usepackage{slashed}
\usepackage{graphicx,amssymb,amsmath,amsfonts}
\usepackage{hyperref}
\usepackage[utf8]{inputenc}

\def\be{\begin{equation}}
\def\ee{\end{equation}}

\def\bmat{\begin{pmatrix}}
\def\emat{\end{pmatrix}}

\def\bdet{\begin{vmatrix}}
\def\edet{\end{vmatrix}}

\numberwithin{equation}{section}
\def\bea{\begin{eqnarray}}
\def\eea{\end{eqnarray}}

\newcommand{\dprime}{{\prime\prime}}
\newcommand{\alp}{\ensuremath{\alpha^\prime}}

\newcommand{\sgn}{\mathrm{sign}}

\newcommand{\pa}{\partial}

\newcommand{\vac}{\vert 0 \rangle}

\newcommand{\nrmO}[1]{\mathop{:}\nolimits\!#1\!\mathop{:}\nolimits}
\newcommand{\nrmB}[1]{\boldsymbol{\mathop{:}}#1\boldsymbol{\mathop{:}}}
\newcommand{\si}{\sigma}

\newcommand{\ep}{\epsilon}

\newcommand{\f}{\frac}

\newcommand{\pil}{\frac\pi\ell}

\title{The scattering amplitude of stringy hadrons I:\\ Strings with opposite charges on their endpoints}

\author{Jacob Sonnenschein and Dorin Weissman and Shimon Yankielowicz}

\affiliation{The Raymond and Beverly Sackler School of Physics and Astronomy,\\
	Tel Aviv University, Ramat Aviv 69978, Israel}

\emailAdd{cobi@post.tau.ac.il}
\emailAdd{dorinw@mail.tau.ac.il}
\emailAdd{shimonya@tauex.tau.ac.il}

\abstract{
In this note we  describe hadrons: mesons and  baryons as strings with electric charges on their endpoints. We consider here only the neutral system with opposite charges, coupled  to an external constant electromagnetic field. We derive certain classical solutions including rotating folded open strings. We write down  the  mode expansion and canonically quantize the system. The OPEs associated with such strings are determined.  We re-derive the non-commutativity of the zero modes and the  fact that the charges modify the spacetime metric. We show that the quantum worldsheet  energy momentum tensor on the boundary is affected  by the endpoint charges and differs from the corresponding  Noether current. 
We determine the generalization of the Veneziano  scattering amplitude for such  strings   in the critical dimension.  Phenomenological implications are addressed and in  particular we show that the external magnetic field can be tuned so that  the amplitude vanishes for  particular kinematic setups.   We  discuss  the generalization of such strings  to  non-critical four dimensional spacetime. In particular we renormalize the divergence of the Polchinski-Strominger effective action associated with the rotating folded string.
}

\begin{document}

\maketitle
\tableofcontents
 
\section{Introduction} \label{sec:intro}
Open strings with electric charges on their ends have been investigated intensively  starting from the early days of string theory and until these days. Many papers have been written about such  open strings interacting with a constant  electromagnetic background fields including several landmark papers \cite{Fradkin:1985qd,Abouelsaood:1986gd,Bachas:1992bh,Seiberg:1999vs}. Reviews of the topic containing more references include \cite{Angelantonj:2002ct,Ambjorn:2000yr}.

What  is then  the reason  to revisit this well studied system?  In the present paper we address this system from a different perspective. While string theory was initially constructed to model hadron physics in the early days of the field, the focus in string theory has shifted away from hadrons by the time the studies cited above were performed. In recent years, the stringy nature of hadrons has been described in the context of the HISH (holography inspired stringy hadron) model \cite{Sonnenschein:2016pim}, with the scope of constructing a phenomenological model of the QCD string. The basic idea here is that hadrons, both mesons and baryons, have the structure of an open bosonic string with
particles on its ends. These particles are massive,  charged and carry spin. Thus, the behavior of  such strings in a constant electromagnetic field   relates to that of, for instance, nucleons or  $\pi$ mesons, propagating in such a background.  
     Unlike  in the usual bosonic string theory,  in which the modes of open strings     correspond  to  fields of  the standard model or other QFTs, here  we associate  them  with the states of  hadrons. The difference between these two points of view  is manifested, for instance, in the fact that in nature the hadronic strings have massive endpoint particles and  an intercept $a\leq 0$\footnote{ Here we mean the intercept of modified Regge trajectories expressed in  terms of the mass squared  as a function of the orbital rather than the total angular momentum. For a discussion of this property see section 3.3 in  \cite{Sonnenschein:2017ylo}.} and hence the spin zero mode does not correspond to a tachyon but rather to a scalar meson of positive mass, and  the spin one does not correspond to a massless gauge field but to a massive vector hadron. This string is also non-critical string theory, since we construct it directly in four dimensions rather than in the critical dimension \(D=26\).

The charges  at the endpoints of strings interact with an external electromagnetic background field as well as interacting electromagnetically one with the other. In fact the latter interaction always takes place in a real world setting, but in the case where the external field is stronger one can approximate the behavior by ignoring the fields induced by the charges. Formally, this is done by not including the kinetic \(F^{\mu\nu}F_{\mu\nu}\) term for the EM field in the action.

It turns out that there is a substantial difference between the case of the interaction with an external EM background field of a string with opposite charges on its ends, such that the string as a whole is neutral, $q_1+q_2=0$, and the case where the charges are general, $q_1\neq -q_2$, and the total charge is non-zero. In the present paper we analyze the neutral string case and the charged string will be discussed in a sequel paper \cite{Sonnenschein:SACSgeneral}.  Strings with charged endpoints with mutual interaction  but with no background electromagnetic field will be addressed in an additional publication.
% \cite{Blech:chargedstrings}. 

The endpoint charges affect  the behavior of the strings in several different aspects: (i) the solutions of the equations of motion, (ii) the classical Regge-like trajectories, (iii) the quantum spectrum and the intercept, (iv) the operator product expansions, (v) the form factors, and (vi) the string scattering amplitudes.  In this paper we analyze each of these aspects for strings with opposite charges.

Let us briefly summarize   the main results of this paper. 
\begin{itemize}
\item
In a constant magnetic field the classical solutions that take the form of a rotating open string with \(n\) folding points, starting from \(n=1\). There is no rotating solution without a fold. These strings admit classical linear Regge trajectories with a tension which is $n T$, where $T$ is the basic string tension. 
The folding points rotate with the speed of light while the speed of the endpoint charges is  $\beta = |\cos\phi|$ where $\phi = \arctan(q B)$.
\item
In previous investigations of this type of strings it has been realized \cite{Abouelsaood:1986gd} that the effect of the interaction with the external EM field is a modification of the spacetime metric,
\be\label{modmetric}
\eta^{\mu\nu}\rightarrow g^{\mu\nu}  = \left(\frac{1}{1-q^2F^2}\right)^{\mu\nu} 
\ee
where $q$ is the charge and $F^{\mu\nu}$ is the background EM field. 
\item
It was further understood that the zero modes of the string, the center of mass coordinates, do not commute 
\be   [x^\mu,x^\nu] = i\theta^{\mu\nu} =i \frac{q}{T} F^\mu{}_\rho g^{\rho\nu} \ee
so that the string coupled to an external field gives a realization of a non-commutative geometry \cite{Seiberg:1999vs}.

\item
The modified spacetime metric enters Veneziano's scattering amplitude via Mandelstam variables that have to be computed using the modified metric. In addition each of the three beta functions is multiplied by a phase associated with the non-commutativity parameters (see eq. \ref{VenAmp}). The phases are a function of the momenta in the directions affected by the field.

\item
En route to the determination of the vertex operators and scattering amplitude we have found that the quantum energy-momentum tensor on the boundary of the worldsheet has a different form compared with the one derived as a Noether current associated with worldsheet translations or by variation with respect to the worldsheet metric. The modification which involves the metric of \ref{modmetric} takes the form 
\be T(y) = -\frac1{2\alp} (g^{-1})_{\mu\nu}\nrmB{\pa_y X^\mu \pa_y X^\nu(y)} \ee
where $y$ is the coordinate on the boundary (the real axis when the worldsheet is the upper half-plane).  It should be emphasized that the bulk energy momentum tensor is not modified and it is only the one on the boundary of the worldsheet. 

\item
We show that for fixed target scattering there are kinematic parameters, namely momenta and scattering angles,  for which by tuning  the magnetic field the scattering amplitude vanishes. 

\item
Since we are interested in the interaction of real neutral hadrons like neutrons or $\rho^0$ with electromagnetic background and their scattering amplitude we have to consider the strings in the non-critical four dimensions. We address this following \cite{Hellerman:2017upi} using the effective string theory approach of Polchinski and Strominger \cite{Polchinski:1991ax}.

\item In the effective string theory approach one expands around a classical solution that has a string length much larger than the string scale. We show how the long string expansion around the rotating string reproduces the same spectrum as the critical theory. To do so we show how one can deal with divergences at folding points, at least one of which is always present for the rotating string in a magnetic field.
\end{itemize}
%%%%%%%%%%%%%%%%%%%%%%%%%%%%%%%%%%%%%%%%% 
%%%%%%%%%%%%%%%%%%%%%%%%%%%%%%%%%%%%%%%%%%%%%%
Our results are in accordance with and complement several previous investigations of the system in particular: the dynamics of open strings in EM background \cite{Nesterenko:1989pz}, spacetime non-commutativity \cite{Seiberg:2000ms}, non-commutative SYM \cite{Russo:2000zb}, holographic description \cite{Hashimoto:1999ut}, a saddle point type of determination of the scattering amplitude \cite{Sever:2009xu} and others.  

The paper is organized as follows: In the next subsection we describe hadrons, both mesons and baryons, as  charged strings in holography and HISH. In section \ref{sec:action} we write down the various forms one can have for the action and the corresponding equations of motion.  In section \ref{sec:symmetries} we write symmetries and conserved currents and charges of the system. In section \ref{sec:modeexp} we present the general solutions of the equations of motion and its mode expansion.
In the next section \ref{sec:classical} we write down explicitly some classical solutions of the equations of motion, in particular a rotating folded string in magnetic field with or without massive endpoints and stretched strings in electric field. In section \ref{sec:canonical} we canonically quantize the system. We determine the non-commutative nature of the zero mode coordinates and we determine the spectrum and compute the intercept of the system.

In section \ref{sec:OPE} we work out the operator product expansion of  the target spacetime coordinates of the string both in the bulk and on the boundary of the worldsheet.  Section \ref{sec:EMtensor} is devoted to the determination of the vertex operator. For that purpose we work out the quantum energy momentum tensor on the boundary. Using this vertex operator we compute in section \ref{sec:scattering} the scattering  amplitude. We show that it takes the form of a modified Veneziano amplitude that incorporates the modified spacetime metric and the non-commutativity parameters.  We also determine certain phenomenological implications of the scattering amplitude.

In section \ref{sec:noncritical} we discuss the system in non-critical dimensions. We invoke the Polchinski-Strominger effective action to determine the intercept of the system and comment on the scattering amplitude in four dimensions. We summarize the results of this paper and present some open questions in section \ref{sec:summary}.

%we describe certain solution of the classical equations of motion.
%In the next section we analyze the symmetries of the system and the corresponding Noether currents and charges. $\S 5$ is devoted to the mode expansions both for the charged and chargeless cases and in particular for a constant magnetic and electric fields. We then in section $\S 6$ canonically quantized the modes for the two cases and two different background fields.
%In the following section we analyze the zero modes and the corresponding non-commutative geometry. In section $\S 8$ we compute the spectra of the different systems and the corresponding intercept. The operator product expansion for the chargeless and the general charged case are computed in section $\S 9$. $\S 10$ is devoted to a brief discussion of the form factors of the system and in  $\S 11$ we determine the modified Veneziano amplitude for the scattering of charged ``tachyonic" strings. In $\S 12$ we summarize our results and mention several open questions. 

%%%%%%%%%%%%%%%%%%%%%%%
\subsection{ Charged strings in holography and HISH}
In nature mesons carry electric charge $+1$, $ 0$, or  $-1$ and baryons $ +2,\  +1,\  0,\  -1$. In holography mesons are strings that end on flavor branes. From the point of view of the flavor branes the ends of the strings carry electric charge and since the string has a different orientation on its two endpoints,  one can refer to one end as a ``quark"  and the other as an ``antiquark".
Baryons in holography are more complicated constructions of strings. They are built from  $N_c$ strings that come out of a $Dp$ brane that wraps a p-cycle and end on flavor branes. 

The HISH model \cite{Sonnenschein:2016pim} maps the holographic stringy hadrons into string configurations in four flat spacetime dimensions and with $N_c=3$.
Instead of strings in curved ten dimensions, the main players in the HISH model  are open strings with  particles on their ends in flat four dimensional spacetime. These particles are massive and carry electric charge and spin. For mesons the particles represent both the ``vertical segments" of the holographic strings and the charge and spin of their end on a flavor brane. This is also the case for one end of the baryonic string and on the other side the endpoint particle is constructed from a baryonic vertex combined with two short strings that connect the baryonic vertex to flavor branes, forming the holographic realization of a diquark. Both in the holographic picture and in the HISH model there is a force that is exerted on the endpoint particles and balances the tension. For hadrons with orbital angular momentum there is a classical centrifugal force that does this. In addition, there is always a repulsive Casimir force \cite{Sonnenschein:2017ylo}, which keeps also hadrons with vanishing orbital momentum from collapsing to a point.

\begin{figure}[ht]
\begin{center}
\includegraphics[width=0.96\textwidth]{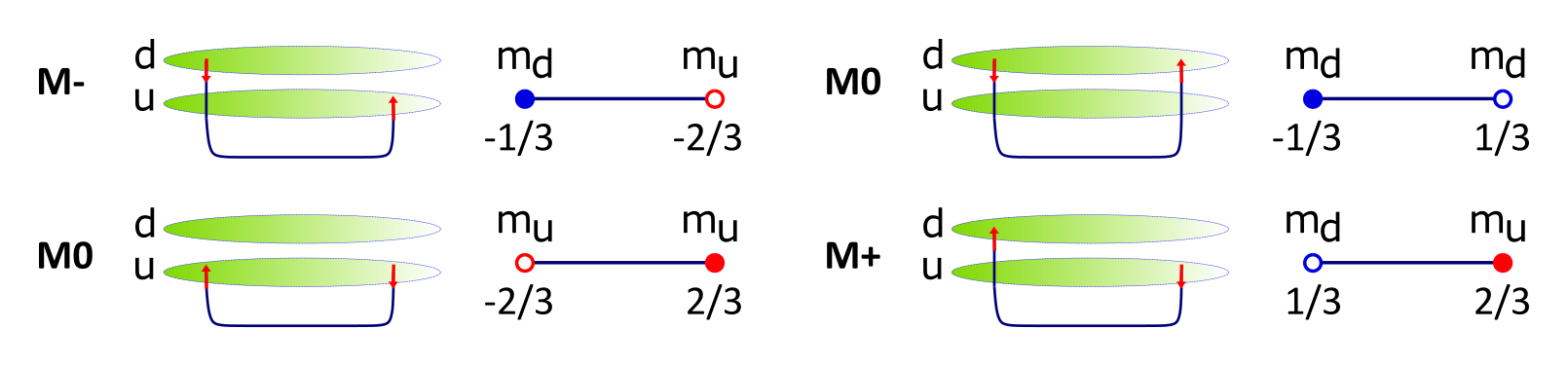}
  \caption{\label{Holcharges} The holographic stringy configurations for mesons with charges $-1,0,+1$ associated with the $u$ and $d$ flavor branes (left) and the corresponding HISH mesons (right).}
 \end{center}
\end{figure}

\begin{figure}[ht]
\begin{center}
\includegraphics[width=0.96\textwidth]{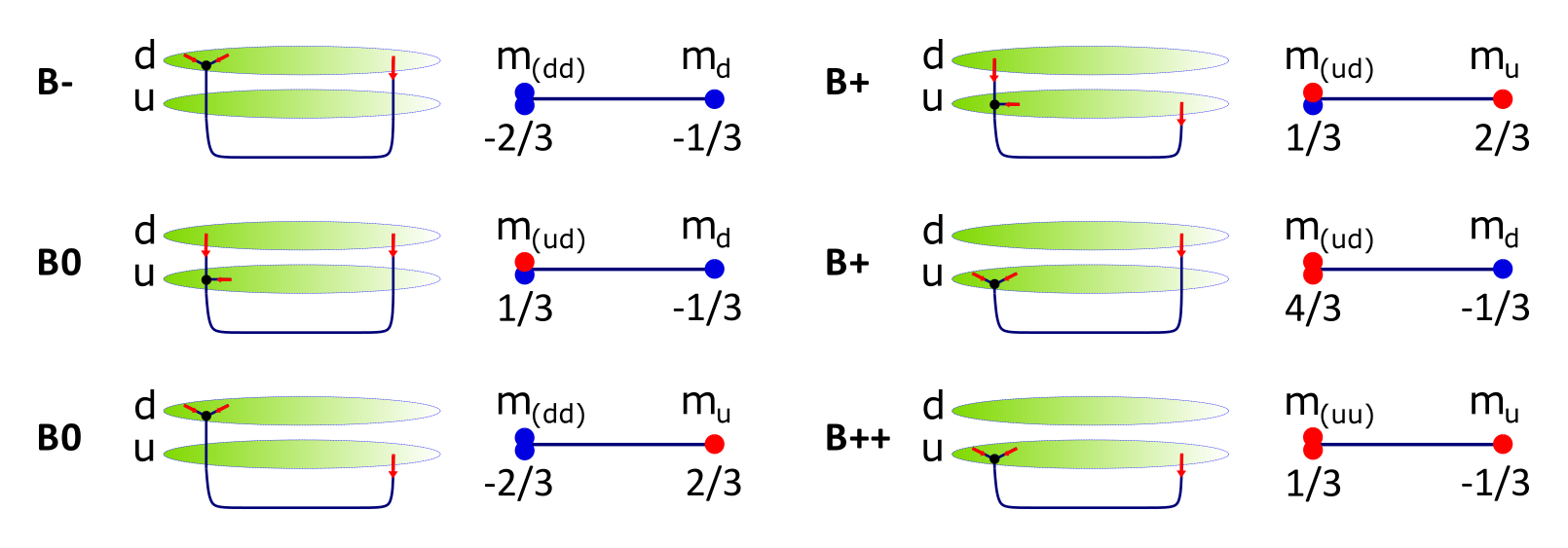}
  \caption{\label{Holbaryons} The holographic stringy configurations for baryons with charges $-1,0,+1,+2$ associated with the $u$ and $d$ flavor branes (left) and the corresponding HISH mesons (right). The diquarks are constructed from a baryonic vertex and two short strings, but are mapped to endpoint particles like the quarks.}
 \end{center}
\end{figure}

We can assign a basic charge to the holographic flavor branes which is the electric charge of the corresponding flavored quark,  namely, 
\[ \frac23 \rightarrow  u\ \mathrm{brane} \qquad -\frac13 \rightarrow  d\ \mathrm{brane} \qquad -\frac13 \rightarrow  s\ \mathrm{brane} \qquad \frac23 \rightarrow  c\ \mathrm{brane} \qquad -\frac13 \rightarrow  b\ \mathrm{brane} \]
then we get for the holographic and HISH hadrons the  charge setups depicted in figures \ref{Holcharges} and \ref{Holbaryons}. The diquarks used to construct the baryon in the HISH picture are endpoint particles like the quarks, and we can associate with them a mass, an electromagnetic charge, and spin.

In figure \ref{Holcharges} we drew the mesons associated with $u$ and $d$ quarks in the holographic picture (left) and in the HISH picture (right) and similarly in  figure \ref{Holbaryons} the corresponding baryons. The vertical direction stands for the holographic radial dimension and the horizontal one for a coordinate in the four dimensional flat space. To describe the HISH string we write an action in the flat spacetime, taking for the string the Nambu-Goto action, plus correction terms from the non-critical theory for dimensions less than 26, as well as appropriate terms on the boundary of the worldsheet for the endpoint particles.

In the current paper we address only the case of neutral hadrons, namely the cases $M0$ and $B0$ in the figures. In a similar manner to the stringy configurations depicted in the figures,  there are stringy  mesons and baryons  that have ends on the $s,c$ and $b$ flavor branes  and their charge setup in built in the same way.

%%%%%%%%%%%%%%%%%%%%%%%%%%%
\section{Action, equations of motion} \label{sec:action}
%%%%%%%%%%%%%%%%%%%%%%%%%%
The action of the system we are interested in is given by
\be\label{action}
 S = S_{st} + \left(S_{pm}+ S_{pq}\right )\vert_{\sigma=0} + \left( S_{pm}+ S_{pq}\right )\vert_{\sigma=\ell} . 
\ee
where $S_{st} $ is the string worldsheet action given by the Nambu-Goto action,
\be
 S_{st} = -T\int d\tau d\sigma \sqrt{-h} = -T\int d\tau d\sigma \sqrt{\dot X^2 X^\prime{}^2 - (\dot X\cdot X^\prime)^2}. 
\ee
In the string action \(h_{\alpha\beta} = \eta_{\mu\nu}\pa_\alpha X^\mu \pa_\beta X^\nu\) is the induced metric on the worldsheet and \(h\) is its determinant, the indices \(\alpha\), \(\beta\) being either \(\tau\) or \(\sigma\). The target space is $D$ dimensional $\mu,\nu= 0,...D-1$. To keep track of units, we take the worldsheet coordinates \(\tau\) and \(\sigma\) to have dimension of length, and our action is defined on the strip: \(-\infty<\tau<\infty\) and \(0\leq\sigma\leq\ell\).

The particles located at both endpoints of the string $\si=0$ and $\si=\ell$ have an action that is built from two parts. First is a mass term $S_{pm}$ given by\footnote{We assume, as in preceding works, that the endpoint particle is spinless.}
\be\label{massaction}
S_{pm} = m_i\int d\tau \sqrt{-\dot X^2}  
 \ee
The second term in the boundary action $S_{pq}$ is
\be\label{chargeaction} 
S_{pq}= T q_i \int d\tau A_\mu(X) \dot X^\mu 
\ee 
The parameters $m_i$ and $q_i$ are the mass and charge of the endpoint particles. We put a factor of the string tension \(T\) before the charged particle action, so that we work in a convention where \(q_i F_{\mu\nu}\) is a  dimensionless quantity, or in other words the electromagnetic fields are given in units of the tension.

At this point we can consider two different cases. In one case we take the action \ref{action} and we treat the gauge field $A_\mu$ as a given background field $A^{back}_\mu(X)$. Another possibility is also to make the gauge field dynamical by turning on a kinetic term, namely taking
\be S \to S -\frac1{4g^2}\int d^4 x F_{\mu\nu}F^{\mu\nu} \ee
In this case the gauge field in $S_{pq}$ takes the form of $A_\mu(X) =A^{back}_\mu(X)+A^{dyn}_\mu(X)$ where $A_\mu^{dyn}$ is the gauge field that one charge induces on the location of the other charge and vice versa, and \(A^{back}\) denotes a background field. In this paper we consider only the former case where the gauge field is a background field. In a sequel paper we will consider the dynamical case. In section \ref{sec:pheno} we write the condition for when neglecting the self interaction is a good approximation for phenomenology.
 
For the case of interest in this paper, of a neutral string with $q_1=-q_2=q$, the action boundary terms  $S_{pq}$ can be rewritten as a bulk term.
% provided certain initial and final conditions are met. 
\be\label{sbu}
S_{sq} =-\f{T}{2}\int d\tau d\sigma \biggl(q F_{\mu\nu} \ep^{\alpha\beta} \pa_\alpha X^\mu \pa_\beta X^\nu \biggr)
\ee
Here \(\ep^{\alpha\beta}\) is the two dimensional antisymmetric tensor with $\ep^{\tau\si }=1$. The bulk action is recognizable as the action for a string in a background of a Kalb-Ramond \(B\)-field.
%The equivalence can be shown as follows 
%\bea
%\frac12\int d\tau d\sigma  F_{\mu\nu} \epsilon^{\alpha \beta} \pa_\alpha X^\mu \pa_\beta X^\nu  & = &   \int d\tau d\sigma \epsilon^{\alpha \beta}\pa_\mu A_\nu \pa_\alpha X^\mu \pa_\beta X^\nu\CR
 % =    2 \int d\tau d\sigma \ep^{\alpha \beta}\pa_\alpha A^\nu \pa_\beta X^\nu & = &   \int d\tau d\si \pa_\alpha(\ep^{\alpha \beta}A^\nu \pa_\beta X^\nu)\CR
 %=  -\int d\tau A_\mu \pa_\tau X^\mu\big|_{\si=\ell} + \int d\tau A_\mu \pa_\tau X^\mu\big|_{\si=0} & + &\int d\sigma A_\mu \pa_\sigma X^\mu\big|_{\tau=+\infty} - \int d\sigma A_\mu \pa_\sigma X^\mu\big|_{\tau=-\infty}\,, \CR \eea 
%Thus, if the sum of the two last terms vanishes the bulk action indeed equals that of the action with the two boundary terms.

%%%%%%%%%%%%%%%%%%%%%%%%%%
%Our starting point is the action defined on the strip, \(-\infty<\tau<\infty\), \(\sigma \in [0,\ell]\),
%\be S\label{action} = -\frac{T}{2}\int d\tau d\sigma (X^{\prime2}-\dot X^2) + T q_1 \int d\tau A_\mu \dot X^\mu\vert_{\sigma=0} + T q_2\int d\tau A_\mu \dot X^\mu\vert_{\sigma=\ell} \ee
From the full action \ref{action} we derive the bulk equations of motion
\be \pa_\alpha(\sqrt{-h}h^{\alpha\beta}\pa_\beta X^\mu) = 0 \ee
and at the boundaries the conditions
\be T\sqrt{-h}\pa^\sigma X^\mu = \pm (m_i\pa_\tau\frac{\dot X^\mu}{\sqrt{-\dot X^2}} + T q_i F^\mu{}_\nu \dot X^\nu) \ee
In the most general case \(F_{\mu\nu}\) is a function of \(X^\mu(\tau,\sigma)\) which is evaluated at the boundary.

The action \ref{action} is invariant under two dimensional reparametrization. We use the symmetry to fix the orthogonal gauge \ref{orthogonal}. In this gauge the bulk equations of motion read
\be X^{\dprime\mu} - \ddot X^\mu = 0 \label{eq:bulkeom} \ee
and the boundary conditions
\be T X^{\prime\mu} +  m_1\pa_\tau\frac{\dot X^\mu}{\sqrt{-\dot X^2}}+ T q_1 F^\mu{}_\nu \dot X^\nu = 0 \qquad \sigma = 0 \label{eq:bd0}\ee
\be T X^{\prime\mu}- m_2\pa_\tau\frac{\dot X^\mu}{\sqrt{-\dot X^2}} - T q_2 F^\mu{}_\nu \dot X^\nu = 0 \qquad \sigma = \ell \label{eq:bdl}\ee
%%%%%%%%%%%%%
Apart from the analysis of classical solutions, in the rest of the paper we will consider only the case where the endpoints are massless, setting $m_1=m_2=0$. For the neutral string case \(q_1+q_2 = 0\) and without masses the boundary equations at $\si=0$ and $\si=\ell$ are the same
\be X^{\prime\mu} + q F^\mu{}_\nu \dot X^\nu = 0 \qquad \sigma = 0,\ \ell \ee
%%%%%%%%%%%%%%%%%%%%%%%%%%%%%%%%%%%%%%%%%%%%%%%%%%%%%%%%%%
\section{ Symmetries and conserved currents} \label{sec:symmetries}
The classical symmetries  of ordinary open string theory are two dimensional diffeomorphism,  local Weyl invariance and target space Poincar\' e invariance. 

We now summarize the symmetries of the combined action of an open string and endpoint charges \ref{action}.

\subsection{Reparametrization invariance and the energy-momentum tensor}
The string action $S_{st}$  is classically invariant under two dimensional local  reparametrization transformations $\{\tau,\sigma\}\rightarrow \{\tilde\tau(\tau, \sigma),\tilde\sigma(\tau, \sigma)\}$, while the boundary action terms $S_{pq}$ and $S_{pm}$ are invariant under one dimensional local reparametrization  $\tau\rightarrow \tilde\tau(\tau)$. It follows that the full action is invariant under reparametrizations that do not move the boundary points \(\sigma = 0\) and \(\ell\).

After fixing the orthogonal gauge,
\begin{align} &\frac12(h_{\tau\tau}+h_{\sigma\sigma})  = \frac12(\dot X^2+X^{\prime2}) = 0 \qquad h_{\tau\sigma}  = h_{\sigma\tau} = \dot X \cdot X^\prime = 0 \label{orthogonal}\end{align}
%The bulk action fixed in this gauge
%\be S_{st} = \frac{-T}{2}\int d\tau d\sigma \left (X^{\prime2}-\dot X^2 \right)\ee
The resulting worldsheet action is still invariant under global translations $\sigma_\alpha\rightarrow \sigma_\alpha + a_\alpha$ where $a_\alpha$ are constants independent of $\tau$ and $\sigma$. The corresponding conserved Noether current is the energy-momentum tensor which reads
\be
T_{\alpha\beta}= \pa_\alpha X^\mu \pa_\beta X_\mu -\eta_{\alpha\beta} {\cal L}
\ee
Classically there is no modification to the energy-momentum tensor from the boundary action.\footnote{One way to see it is that the boundary action will not depend on the einbein on the endpoint's worldline \cite{Fradkin:1985qd}, so when writing \(T_{ab}\) as the variation of the action under an independent worldsheet metric in the Polyakov formulation, there is no contribution from the boudnary.} That is the classical result. It will be shown later in section \ref{sec:EMtensor} that quantum mechanically the energy-momentum tensor is modified by the boundary charge terms.
\subsection{Weyl invariance} 
%It is straightforward to show that the theory is classically scale invariant. 
%The boundary action (\ref{chargeaction}) and the bulk action (\ref{sbu}) do not depend on the worldsheet metric and hence they are both trivially Weyl invariant classically. As was just mentioned the boundary mass term $S_{pm}$ does depend on worldline einbein $e(\tau)$ but not $S_{pq}$.  

Quantum mechanically Weyl invariance of the boundary action requires the vanishing of the beta function associated with the coupling to the gauge field. It was shown in \cite{Abouelsaood:1986gd} that the condition for the vanishing of the  beta function is given by
\be
\pa^\nu F_{\mu}{}^\lambda \left(\frac{1}{1-q^2 F^2}\right )_{\lambda \nu} =0
\ee
which to the leading order in $\alp$ is Maxwell's equation (for the derivation of the matrix factor see \ref{matrixg}). Constant $F_{\mu\nu}$ which will be assumed in the rest of the paper is a trivial solution of the above equation. 

\subsection{Target space Poincar\'e invariance}
Assuming a constant background field \(F_{\mu\nu}\) the action is invariant under global spacetime translations
\be
\delta X^\mu(\sigma,\tau) = \epsilon^\mu
\ee
for constant \(\epsilon^\mu\).
%since its   dependence on $X^\mu$ is only via its derivatives $\pa_\tau X^\mu$ and $\pa_\sigma X^\mu$. However, this is not the case for the action that includes boundary terms (\ref{chargeaction}). In this case the variation of the action under the translation reads
%\be\label{varsptran}
%\delta S = q \epsilon^\lambda \left[\int d\tau \pa_\lambda A_\mu(X)  \dot X^\mu|_{\sigma=0} - \int d\tau \pa_\lambda A_\mu(X)  \dot X^\mu|_{\sigma=\ell} \right ] 
%\ee
%For a constant EM field we can pick the gauge where $A_\mu =-\frac12 F_{\mu\nu} X^\nu$. For a purely electric or purely magnetic field, we can choose coordinates so that $F_{\mu\nu}$ is non-vanishing along a single pair of coordinates, and then translations along the two other coordinates do not modify the boundary terms. In any case, translations along the directions of $F_{\mu\nu}$ do modify the boundary terms.
%The variation  of these terms takes the form
%\be
%\delta S = \lim_{T\rightarrow \infty} q F_{\mu\lambda}\epsilon^\lambda
%\left[  X^\mu(T,0)- X^\mu(-T,0) -X^\mu(T,\ell) + X^\mu(-T,\ell)\right]
%\ee
%When the square bracket  vanishes the action is invariant also  under spacetime translations along the direction of the EM field. This condition is related to the condition of the equivalence of the bulk and boundary terms formulations. 

%For  the action which  is invariant under spacetime translations, either the bulk action or the the one that includes boundary terms and the condition of above is fulfilled, 
The corresponding Noether charge \(P^\mu\) is given by
\be P_\mu = T\int d\sigma \dot X_\mu - T q F_{\mu\nu} X^\nu|_{\sigma=0} + T q F_{\mu\nu}X^\nu|_{\sigma=\ell} \label{eq:pmu} \ee
%For components transverse to $F_{\mu\nu}$ only the term which is an integral over $\sigma$ survives, while in other directions we also get the contribution from the boundary.
%%%%%%%%%%%%%%%%%%%%%%%%%%%%%%%%%%%%%%%%%%%%%%%%%%%%%%%
%\subsubsection {Lorentz transformations}
The bulk equation is invariant under Lorentz transformations,
\be \delta X^\mu = \omega^\mu{}_\nu X^\nu, \ee
%where \(\omega_{\mu\nu} = -\omega_{\nu\mu}\) is antisymmetric. The Nambu-Goto action is invariant in a trivial way, while for the Lorentz transformations of the coupling to the EM field the situation is as follows. 
with \(\omega_{\mu\nu}\) antisymmetric. Since \(F_{\mu\nu}\) is a constant background field, the boundary action is invariant under transformations that leave it invariant
\be \delta F_{\mu\nu} = \omega_\mu{}^\rho F_{\rho\nu} - \omega_\nu{}^\rho F_{\rho\mu} = 0 \label{eq:deltaF} \ee
%This can be verified for both the boundary (\ref{chargeaction}) and bulk (\ref{sbu}) actions coupling the string to the EM field.
%For Lorentz transformations to be a symmetry there is no requirement at \(\tau \to \pm\infty\) as with translations.
For any \(F_{\mu\nu}\) in four dimensions there are two independent transformations that obey this constraint (appendix \ref{app:Lorentz}).

Generally, the generators of the Lorentz transformation are
\be J^a_{\mu\nu} = T\left[\sqrt{-h}h^{ab} (X_\nu \pa_b X_\mu - X_\mu \pa_b X_\nu) + q\epsilon^{ab}(F_{\mu\rho} X_\nu \pa_b X^\rho-F_{\nu\rho} X_\mu \pa_b X^\rho)\right] \ee
The Noether charges are, writing the part from the NG action in the orthogonal gauge,
\be J_{\mu\nu} = T\int d\sigma\left[X_\mu \dot X_\nu - X_\nu \dot X_\mu +  q F_{\mu\rho} X_\nu \pa_\sigma X^\rho - q F_{\nu\rho} X_\mu \pa_\sigma X^\rho\right] \ee
Of course, only the charges (or combinations of charges) that correspond to transformations leaving \(F_{\mu\nu}\) invariant are conserved.

\subsection{Gauge invariance}
%The worldsheet action in its bulk formulation (\ref{sbu}) is obviously gauge invariant, since it is written in terms of the field strength \(F\).  Under the gauge transformation $ A_\mu(X)\rightarrow A_\mu(X) +\pa_\mu \Lambda(X) $ , the variation of the boundary term at $\sigma= 0$ takes the form
%\be\label{gaugevar}
%\delta S_{pq} = q_i \int d\tau \pa_\mu \Lambda(X) \dot X^\mu = q_i \int d\tau \pa_\tau \Lambda(X) =  q_i\lim_{T\rightarrow \infty}  
%\left[\Lambda(X(T,0))-  \Lambda(X(-T,0))  \right]
%\ee
%and similarly for the boundary at $\sigma= \ell$. Thus, the action with boundary terms is invariant under gauge transformations only provided that the transformation parameter is periodic or  vanishes on the boundary of the $\tau$ direction.

The invariance of the classical action under gauge transformation $A_\mu(X)\rightarrow A_\mu(X) +\pa_\mu \Lambda(X)$ is easy to prove. As was discussed in detail in \cite{Seiberg:1999vs} in the quantum picture the system has to be regularized. Using point splitting regularization and taking into account the non-commutative nature of the zero modes (see eq. \ref{eq:ncg}) it was shown that the quantum action is in fact invariant under a modified gauge transformation, involving the Moyal star product defined using the non-commutativity phase.

%In (\ref{varsptran}) it was pointed out that invariance under spacetime translations requires fulfilling certain boundary conditions. Similarly (\ref{gaugevar}) sets the condition for invariance under gauge transformation. It is easy to check that the action is invariant under combined spacetime translation with \(a^\mu\) and gauge transformation with \(\Lambda(X)\), if the transformation parameters obey $ a^\nu\pa_\nu A_\mu -\pa_\mu \Lambda(X)=0$ at the endpoints  $\sigma=0$ and $\sigma=\ell$.
%\be
%\delta A_\mu = \pa_\mu \Lambda(X) + i \Lambda(X) \odot A_\mu(X) - i A_\mu(X)\odot \Lambda(X)
%\ee  
%where the $\odot$ product will be  defined in (\ref{star}).
%%%%%%%%%%%%%%%%%%%%%%%%%%%%%%%

%\subsection{Energy momentum tensor}
%In the case of \(q_1 = -q_2 = q\) and a constant \(F_{\mu\nu}\), we take the gauge \(A_\mu = -\frac12 F_{\mu\nu}X^\nu\) and then we can write the action as
%\be S = -\frac{T}{2}\int d\tau d\sigma (X^{\prime2}-\dot X^2) + \frac 2 T q \int d\tau F_{\mu\nu} \dot X^\mu\vert_{\sigma=0}^\ell \ee
%Now we can convert the boundary term to a bulk term,
%\be S = -T\int d\tau d\sigma (\frac12X^{\prime2}-\frac12\dot X^2 - q F_{\mu\nu} \dot X^\mu X^{\prime\nu}) \ee
%or
%\be S = -\frac{T}{2}\int d\tau d\sigma (\eta_{\mu\nu}\eta^{ab}- q F_{\mu\nu}\epsilon^{ab})\pa_a X^\mu \pa_b X^\nu \ee
%This is the action of the string in a background with a non-vanishing Kalb Ramond \(B\)-field.

\section{General solution of the equations of motion} \label{sec:modeexp}
The most general solution of the bulk equations of motion (\ref{eq:bulkeom}) is a sum of left and right moving modes,
\be X^\mu(\tau,\sigma) = X_R^\mu(\tau-\sigma) + X_L^\mu(\tau+\sigma) \ee
for which can we write the mode expansions
\be X_R^\mu = x_R^\mu + \alpha_0^\mu (\tau-\sigma) + i\sqrt{N}\sum_{n}\frac{\alpha^\mu_{n}}{\omega_n}e^{-i\frac\pi\ell \omega_n(\tau-\sigma)}\ee
\be X_L^\mu = x_L^\mu +\tilde\alpha_0^\mu (\tau+\sigma) + i\sqrt{N}\sum_{n}\frac{\tilde\alpha^\mu_{n}}{\omega_n}e^{-i\frac\pi\ell \omega_n(\tau+\sigma)}\ee
The two boundary conditions at \(\sigma = 0\) and \(\ell\) determine the spectrum of allowed eigenfrequencies \(\omega_n\) and the relation between the left and right moving oscillators, \(\alpha_n\) and \(\tilde\alpha_n\). The normalization constant \(N\) will be determined later to be \(N = \frac1{4\pi T} = \frac\alp2\).

We will write the equations for the general case of any \(q_1\) and \(q_2\) before specializing to the neutral string with \(q_1=-q_2=q\). The boundary conditions we have to satisfy are
\begin{align} X^{\prime\mu} + q_1 F^\mu{}_\nu \dot X^\nu = 0\,, &\qquad \sigma = 0 \\
							X^{\prime\mu} - q_2 F^\mu{}_\nu \dot X^\nu = 0\,, &\qquad \sigma = \ell \end{align}
The first difference between the neutral and charged strings is that the zero modes linear in \(\tau\) and \(\sigma\) cannot satisfy both conditions simultaneously, unless \(q_1 = -q_2\) and the boundary condition is the same on both endpoints.

For the oscillating modes, the two boundary conditions translate into the requirements
\begin{align} (\delta^{\mu}_\nu- q_1 F^{\mu}{}_\nu)\alpha^\nu_n &= (\delta^{\mu}_\nu+ q_1 F^{\mu}{}_\nu)\tilde \alpha^\nu_n \\
 (\delta^\mu_\nu + q_2 F^\mu{}_\nu)\alpha_n^\nu &= e^{-2i\pi\omega_n}(\delta^\mu_\nu - q_2 F^\mu{}_\nu)\tilde\alpha_n^\nu \end{align}

We define the two matrices
\be M_1 = (1+q_1 F)^{-1}(1-q_1 F) \ee
\be M_2 = (1+q_2 F)^{-1}(1-q_2 F) \ee
Such that the first boundary condition can be written in matrix notation\footnote{Throughout this paper we use matrix notation to denote tensors where the first index is an upper index and the second is a lower index. When we define, for instance, \(C = (1-q F)^{-1}\) we mean the tensor \(C^\mu{}_\nu\) that obeys \(C^\mu{}_\alpha (\delta^\alpha_\nu - q F^\alpha{}_\nu) = \delta^\mu_\nu\).}
as \(\tilde \alpha_n = M_1 \alpha_n\), and then the second boundary condition is
\be M_2 M_1 \alpha = e^{2i\pi\omega_n} \alpha \label{eq:eigenvector}\ee
For \(q_1 = -q_2\) then the matrices \(M_1\) and \(M_2\) are the inverse of one another, and we simply get that the solution is \(\omega_n = n\). For the general case the possible solutions are \(\omega_n = n +\delta\), where \(\delta\) is independent of \(n\) and is a simple function of the eigenvalues of the matrix \(M_2 M_1\).  The charged string case will be discussed fully in the sequel paper \cite{Sonnenschein:SACSgeneral}.

Returning to the neutral string case then, denoting \(q = q_1 = -q_2\), we have seen from the above that the left and right moving modes are related by
\be \tilde \alpha^\mu_n = M^\mu{}_\nu \alpha^\nu_n \ee
where the matrix \(M\) is defined as
\be M = (1+qF)^{-1}(1-qF) \label{eq:MF}\ee
The full mode expansion that solves the equations of motion and the boundary conditions in the neutral case is
\be X^\mu(\tau,\sigma) = x^\mu + \alpha_0^\mu \tau - q F^\mu{}_\nu \alpha_0^\nu (\sigma-\frac\ell2) + i\sqrt{N}\sum_{n}\frac{\alpha^\nu_{n}}{n}e^{-i\frac\pi\ell n\tau}\left(e^{i\frac\pi\ell n\sigma}\delta^\mu_\nu + e^{-i\frac\pi\ell n \sigma}M^\mu{}_\nu\right) \label{eq:modeexp}\ee

In the above expression we have written the zero mode such that the parameter \(x^\mu\) is the center of mass coordinate (at \(\tau=0\)),
\be x^\mu = \frac1\ell\int_0^\ell d\sigma X^\mu(\tau,\sigma)|_{\tau=0} \ee
while \(\alpha_0^\mu\) is related to the total momentum of the string by
\be p^\mu = T\int d\tau \dot X^\mu - T q F^\mu{}_{\nu} (X^\nu|_{\sigma=0} - X^\nu|_{\sigma=\ell}) = T\ell (\delta^\mu_\rho - q^2 F^\mu{}_\nu F^\nu{}_\rho)\alpha_0^\rho \ee
Or
\be \alpha_0^\mu = \frac{1}{T\ell}g^\mu{}_\nu p^\nu \qquad g \equiv (1-q^2 F^2)^{-1}\label{matrixg} \ee
We will see that the matrix that we have denoted \(g\) becomes important later on, where we see how it can be interpreted as an effective metric seen by the open string.

Furthermore, the matrix \(M\) relating the left and right moving oscillators \(\alpha_n\) and \(\tilde\alpha_n\) can be seen to be a Lorentz transformation. One can verify that for any choice of \(F_{\mu\nu}\) it obeys
\be \eta^{\alpha\beta}M^\mu{}_\alpha M^\nu{}_\beta = \eta^{\mu\nu} \ee
For a purely magnetic field \(M\) is a rotation, while for an electric field it is a boost. Some explicit examples are in appendix \ref{app:Solutions}.

%\subsection{World sheet parity transformation}
%Our system has the symmetry
%\be \sigma \to \ell -\sigma \qquad q_1 \leftrightarrow q_2 \label{eq:parity}\ee
%That is, the string is symmetric under exchange of the two charges, but the orientation of the string also has to be changed. The transformation above defines the world sheet parity operator \(\Omega\), with \(\Omega^2 = 1\), under which \(X\) should be invariant
%\be \Omega X^\mu(\tau,\sigma) \Omega^{-1} = X^\mu(\tau,\sigma) \ee
%We can apply the transformation of eq. \ref{eq:parity} to the mode expansion:
%\begin{align} \Omega X^\mu(\tau,\sigma)\Omega^{-1} &= \Omega x^\mu \Omega^{-1} + \Omega\alpha_0^\mu\Omega \tau - q F^\mu{}_\nu \Omega\alpha_0^\nu\Omega^{-1} (\frac\ell2-\sigma) + \\ \nonumber &+ i\sqrt{N}\sum_{n}\frac{\Omega\alpha^\nu_{n}\Omega^{-1}}{n}e^{-i\frac\pi\ell n\tau}\left(e^{i\pi n} e^{-i\frac\pi\ell n\sigma}\delta^\mu_\nu + e^{-i\pi n} e^{i\frac\pi\ell n \sigma}(M^{-1})^\mu{}_\nu\right) \label{eq:modeexp}\end{align}
%We have used the fact that exchanging the charges \(q\to-q\) takes \(M\) to \(M^{-1}\). For \(X^\mu\) to be invariant under this transformation, the zero modes should be invariant
%\be \Omega x^\mu \Omega^{-1} = x^\mu \qquad \Omega\alpha_0^\mu \Omega^{-1} = \alpha_0^\mu \ee
%while the \(\alpha^\mu_n\) have to transform as
%\be \Omega \alpha^\mu_n \Omega^{-1} = (-1)^n M^\mu{}_\nu \alpha_n^\nu \ee

%%%%%%%%%%%%%%%%%%%%%%%%%
\section{Classical solutions} \label{sec:classical}
%We start with the solutions of a relativistic particle both for an external electric and magnetic fields.  We then discuss certain special cases of strings classical motion in the external field depending on certain initial motion of the string before turning on the external field.
Before quantizing the modes, we will write explicitly some classical solutions of the equations of motion that are of special interest.

\subsection{Rotating string in magnetic field} \label{sec:rotating}
For a string in a constant magnetic field \(F_{12} = B\), then
\be X^0 = e \tau \qquad X^1 = \frac e \omega \cos(\omega \sigma + \phi)\cos(\omega \tau) \qquad X^2 = \frac e \omega \cos(\omega \sigma + \phi)\sin(\omega \tau) \label{eq:rotsol}\ee
is a solution to the equations of motion and boundary conditions provided
\be \omega = \pil n \qquad \phi = \arctan(q B) \ee
The solution also obeys the Virasoro constraints \(\dot X^2+X^{\prime2}=0\) and \(\dot X\cdot X^\prime=0\). Solutions with \(n>1\) describe strings that are folded on themselves \(n\) times. Under the magnetic field, the solution with \(n=1\) also develops a fold at \(\sigma = \ell(1-\phi/\pi)\) (we will assume w.l.o.g. that \(qB>0\) and hence \(\phi>0\)).

\begin{figure}[h]
\begin{center}
\includegraphics[width=0.48\textwidth]{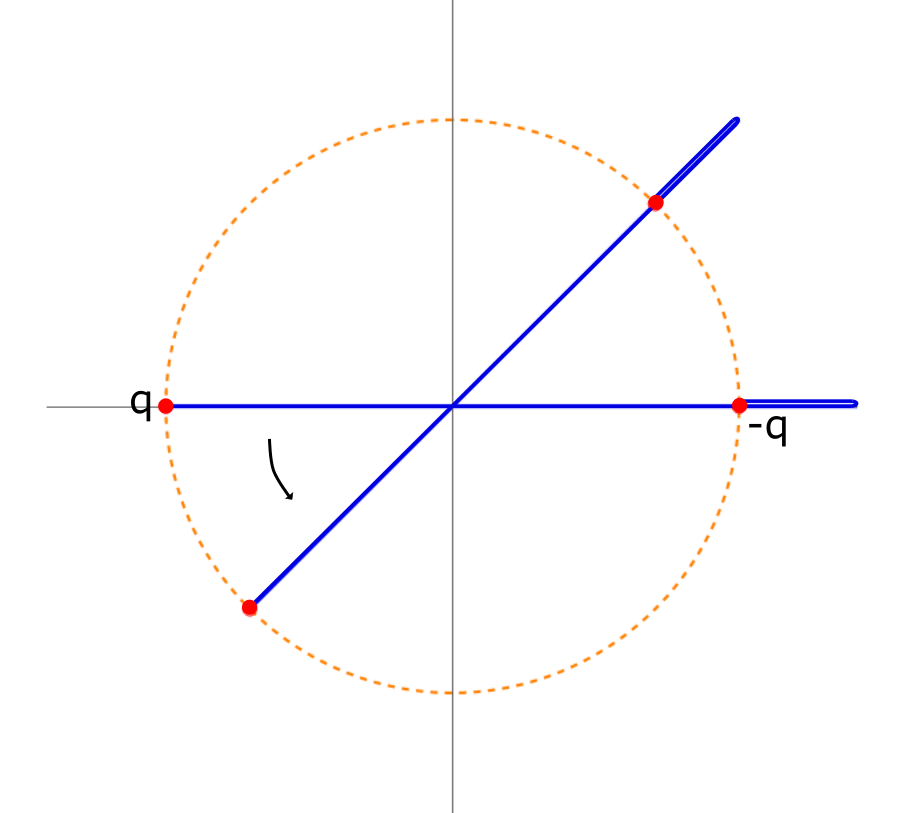}
  \caption{\label{fig:rotating_neutral} The rotating solution of a neutral string with endpoint charges in a magnetic field. The charges rotate around the central point between them. The magnetic field is in the \(z\) direction, coming out of the plane, and the rotation is counterclockwise.}
 \end{center}
\end{figure}

The energy and angular momentum of the rotating string are given by
\be E = T e \ell \ee
\be J = \frac{T e^2 \ell^2}{2\pi n} \ee
Note that the endpoints do not contribute to \(E\) and \(J\).

We can already see that
\be J = \frac{1}{2\pi T n}E^2 \ee
For \(n=1\) this is the usual open string trajectory, while larger \(n\) gives the result for a folded string, where the effective tension is \(n\) times the basic tension \(T\).

The length of the string is
\be L = \int_0^\ell d\sigma |R^\prime(\sigma)| = e\int_0^\ell |\sin(\omega\sigma+\phi)| = \frac2\pi e\ell \ee
So that \(E\) and \(J\) are given by the familiar expressions \(E = \frac\pi2 TL\), \(J = \frac{1}{n}\frac\pi8 TL^2\). Note that \(L\) is the total length of the folded string, which is \(n\) times the extent of the string in space time, plus the additional folded segment whose length depends on the magnetic field.

Points along the string move at a velocity given by
\be \gamma^{-1}(\sigma) = -\frac1e\sqrt{-\dot X^2} = |\sin(\omega\sigma+\phi)| \qquad |\beta(\sigma)| = |\cos(\omega\sigma+\phi)| \ee
Contrary to the rotating open string with Neumann  boundary conditions, here the endpoints move at a finite velocity given by
\be \gamma^{-1} = |\sin\phi| \qquad \beta = |\cos\phi| \ee
while the folding point in the string moves at the speed of light.

\subsection{Rotating solution with endpoint masses}
The classical solutions may be easier to understand when we add mass terms on the endpoints of the string in addition to charges. Then, the boundary condition of the string is actually the force equation on the endpoint particles, and we can see under what conditions the fold develops. The boundary conditions with masses were already stated in eqs. \ref{eq:bd0}-\ref{eq:bdl}.

For the ansatz of a rotating solution as in eq. \ref{eq:rotsol}, the boundary conditions are
\begin{align} &-T e \sin\phi + \frac{m_1\omega \cos\phi}{|\sin\phi|} + e q B \cos\phi = 0 &\qquad \sigma = 0 \nonumber \\  &T e \sin(\omega\ell+\phi) + \frac{m_2\omega \cos(\omega\ell+\phi)}{|\sin(\omega\ell+\phi)|} - e q B \cos(\omega\ell+\phi) = 0 &\qquad \sigma = \ell \label{eq:bdm} \end{align}
When \(m = 0\) we find the solution from before with \(\omega\ell=n\pi\) and \(\phi = \arctan(q B)\).

We want to see now the directions of the forces acting on the particle, and why the fold has to appear. At time \(\tau = 0\), what we wrote above is the force equation on the particle in the \(X^1\) direction.

In the first equation, if \(\phi = \phi_0\) is a solution then so is \(\phi_0+\pi\). We pick the solution between \(-\pi/2\) and \(\pi/2\) so \(\cos\phi\) is positive. We can also choose \(q B>0\) and \(\omega>0\) so that \(\sin\phi\) is also positive. The first equation can be written in terms of the velocity of the first endpoint as
\be \gamma_1 m_1 \omega \beta_1 - \frac{T e}{\gamma_1} + e q B \beta_1 = 0\ee

\begin{figure}[h]
\begin{center}
\includegraphics[width=0.48\textwidth]{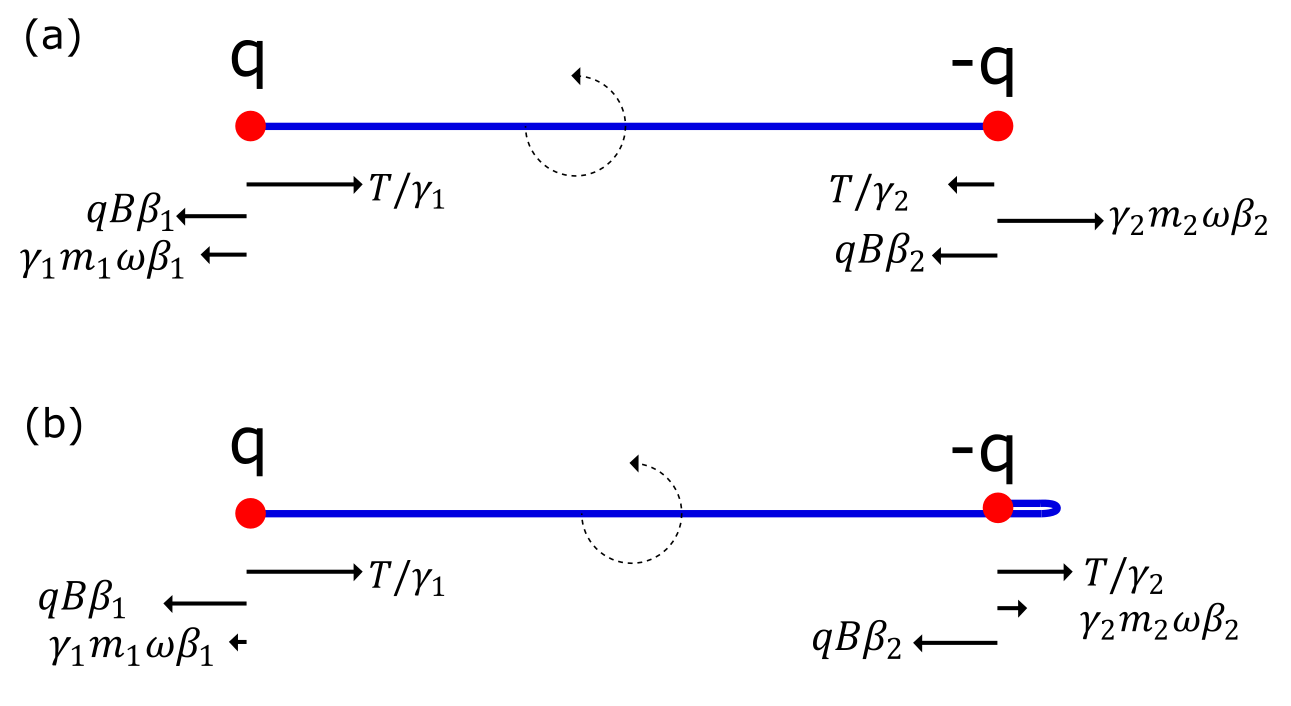}
  \caption{\label{fig:rotating_neutral_fold} The forces acting on the endpoint particles when the fold develops. The forces are drawn to scale for a solution with \(qB=1/8\) and equal endpoint masses. In (a) we take \(m/Te\ell=0.003\) (\(Te\ell\) being equal to the energy carried by the string) and we find a solution without a fold. When the masses are taken to be even smaller, \(m/Te\ell\) = 0.001 then we have only the solution with the fold.}
 \end{center}
\end{figure}

For the other endpoint, we take the first solution, in which \(0<\omega\ell<\pi\). Now there are two possibilities, either the solution requires \(\omega\ell+\phi > \pi\) (as we have seen it is in the massless case), or not. If it is larger than \(\pi\) then a fold develops. With the fold, then the pull of string is added to the centrifugal force, rather than subtracted from it:
\be -\gamma_2 m_2 \omega \beta_2 - \frac{Te}{\gamma_2} + T e q B \beta_2 = 0  \ee
If there is no fold, then the sign of the tension force is reversed, while the others remain the same
\be -\gamma_2 m_2 \omega \beta_2 + \frac{Te}{\gamma_2} + T e q B \beta_2 = 0  \ee
The two options are drawn in figure \ref{fig:rotating_neutral_fold}. Solving the boundary equations \ref{eq:bdm} numerically shows that, for a given value of \(qB\) there is a value of the mass below which there must be a fold in order to obey the force equations.

\subsection{Stretched string in electric field}
The simplest solution for a string stretched by an electric field in a given direction, w.l.o.g. \(F_{01} = -E\) is
\be X^0 = \tau \qquad X^1 = -q E \sigma \qquad X^t = p^t \tau \ee
Where \(X^t\) is some direction transverse to \(X^0\) and \(X^1\). The transverse momentum is needed to make the solution obey the Virasoro constraint \(\dot X^2 + X^{\prime2} = 0\) (or the full Nambu-Goto equations of motion), and it is
\be p_t^2 = 1-q^2 E^2 \ee
We can understand this condition again by introducing an endpoint mass so that the boundary condition is the force equation on the endpoint particle. For the solution above, the boundary condition in the presence of an endpoint mass in the \(X^1\) direction is
\be m\pa_\tau\left(\frac{\dot X^1}{\sqrt{-\dot X^2}}\right) = T\sqrt{1-p_t^2} - T q E \ee
The LHS is zero since the solution is not accelerating. The condition then that the forces be balanced can be met for a specific value of \(p_t^2\), as the tension is rescaled by a factor of \(\sqrt{1-p_t^2} = \gamma^{-1}\) in the force equation.

Alternatively, one could have a rotating solution in a plane transverse to the electric field for the same effect
\be X^0 = \tau \qquad X^1 = -q E \sigma \qquad X^2 = A\cos(\omega\sigma)\cos(\omega\tau) \qquad X^3 = A\cos(\omega\sigma)\sin(\omega\tau) \ee
with
\be \omega = \pil \qquad A^2\omega^2 = 1-q^2 E^2 \ee

%%%%%%%%%%%%%%%%%%%%%%%%%%%%%
\section{Canonical quantization} \label{sec:canonical}
\subsection{Quantizing the modes}
Now we take the mode expansion of section \ref{sec:modeexp} and perform the quantization of the modes. We quantize by imposing the canonical equal time commutation relations
\be [X^\mu(\tau,\sigma),X^\nu(\tau,\sigma^\prime)] = 0 \qquad
 [X^\mu(\tau,\sigma),\Pi^\nu(\tau,\sigma^\prime)] = i\eta^{\mu\nu}\delta(\sigma-\sigma^\prime) \ee
where the congujate momentum to \(X^\mu\) is
\be \Pi_\mu = \frac{\delta\mathcal L}{\delta \dot X^\mu} = T \left(\dot X^\mu + q A_\mu \delta(\sigma) - q A_\mu\delta(\sigma-\ell) \right) \ee
For constant \(F_{\mu\nu}\) we use he gauge \(A_\mu = -\frac12F_{\mu\nu}X^\nu\).

The canonical commutation relations hold provided that
\be [x^\mu,p^\nu] = i\eta^{\mu\nu} \qquad [\alpha_m^\mu,\alpha_n^\nu] = m\eta^{\mu\nu}\delta_{m+n} \qquad [x^\mu,\alpha^\nu_{n\neq0}] = 0 \ee
Note that the canonical \([x^\mu,p^\nu]\) commutator implies that for the mode \(\alpha_0\),
\be [x^\mu,\alpha_0^\nu] = \frac{i}{T\ell}g^{\mu\nu} \ee
where \(g = (1-q^2 F^2)^{-1}\) as defined above. One also finds that for the commutator \([X^\mu(\sigma),X^\nu(\sigma^\prime)]\) to be zero one needs to leave a non-zero commutator of the center of mass coordinates
\be [x^\mu,x^\nu] = i\theta^{\mu\nu} = i\frac{q}{T}F^{\mu}{}_\rho g^{\rho\nu} \label{eq:ncg}\ee
This is the celebrated non-commutative geometry discussed in \cite{Seiberg:1999vs} and followup works. For the specific case of a constant electric field it is a time-space non-commutativity
\be [x^0,x^1] = \frac{i}{T}\frac{q E}{1-q^2 E^2} \ee
while in the constant magnetic field \(F_{12} = B\) the non-trivial commutation relation is
\be [x^1,x^2] = \frac{i}{T}\frac{q B}{1+q^2 B^2} \ee

\subsection{The spectrum and the intercept} \label{sec:spectrum}
%%%%%%%%%%%%%%%%%%%%%%%%%%%%%%%%%%%%%%%%%%%%
The worldsheet Hamiltonian of the system of the string with endpoint charges is
\be H = \frac{T}{2}\int_0^\ell  d\sigma (\dot X^2+ X^{\prime2}) \label{eq:Hws} \ee
with no contribution from the boundary terms. If we insert the mode expansion for \(X\) (eq. \ref{eq:modeexp}) into the above, we find the result
%\be H = \frac{T}{2}\int_0^\ell (\dot X^2 + X^{\prime2}) = \frac{\pi}{\ell} \alp (M_0)_{\mu\nu}p^\mu p^\nu + \frac{\pi}{2\ell} \sum_{n\neq0} \eta_{\mu\nu}\alpha_{-n}^\mu \alpha_n^\nu \ee
%or
\be \frac{\ell}{\pi} H = \alp g_{\mu\nu}p^\mu p^\nu + \frac12\sum_{n\neq0}\eta_{\mu\nu} \alpha_{-n}^\mu \alpha_n^\nu \label{eq:H_chargeless}\ee
with
\be g_{\mu\nu} = (\frac{1}{1-q^2 F^2})_{\mu\nu} \ee
and where \(\alp = (2\pi T)^{-1}\) is the usual Regge slope. Quantum mechanically, the only correction that \(H\) will take will be the addition of the normal ordering constant, which is the same as for the Neumann string since the eigenfrequencies are the same and the regularization procedure can be done without reference to the external field,
\be a = -\frac{D-2}{2}\sum_{n=1}^\infty n = \frac{D-2}{24}  = 1 \ee
To get this result we used the simplest Zeta function regularization of the divergent sum, \(\sum_n n = \zeta(-1) = -1/12\), but it can also be obtained in other ways, for instance by looking at the algebra of the Virasoro generators as was done in \cite{Abouelsaood:1986gd}. In the last equation we set \(D=26\). In section \ref{sec:noncritical} we look at the semiclassical quantization around a rotating string solution, which replicates this result and where the renormalization can be done in a different way. The intercept for the non-critical string will include the same \((D-2)/24\) from the fluctuations plus a contribution from the Polchinski-Strominger term in the effective such that the total intercept is \(a = 1\) for any dimension \(D\) (see section \ref{sec:noncrita}).

The mass-shell condition for a physical state \((L_0-a)\vert \Phi \rangle = 0\) now includes the effective metric, and it is
\be g_{\mu\nu}p^\mu p^\nu = -\frac1\alp(N - a) \ee
On the other hand, the string is still propagating in a flat background metric, so one might define its mass squared as
\be M^2 = -\eta_{\mu\nu}p^\mu p^\nu = \frac1\alp(N-a) + (g_{\mu\nu}-\eta_{\mu\nu})p^\mu p^\nu \ee
or
\be M^2 = \frac1\alp(N-a) + q^2 F_{\mu\alpha}g^{\alpha\beta}F_{\beta\nu}p^\mu p^\nu \ee

The explicit expressions for the magnetic (\(F_{12} = B\)) and electric (\(F_{10}=E\)) fields are, respectively
\be M^2 = \frac1\alp(N-a) - \frac{q^2B^2}{1+q^2B^2}(p_1^2+p_2^2) \ee
\be M^2 = \frac1\alp(N-a) - \frac{q^2E^2}{1-q^2E^2}(p_0^2-p_1^2) \ee

It is not clear what the correct definition of the mass is, whether it is to be defined with the effective metric \(g_{\mu\nu}\) or with the background metric \(\eta_{\mu\nu}\). The Lorentz symmetry is broken by the external field, and both possible definitions of the mass are invariant under the remaining unbroken part of the symmetry.

Some authors addressing this issue in the past have defined it the former way (for instance \cite{Abouelsaood:1986gd}), declaring the spectrum of the string is unchanged by the electromagnetic field, while others used the definition with \(\eta_{\mu\nu}\), noting the negative sign contribution to the mass squared as we have here \cite{Nesterenko:1989pz}. Most references do not relate \(p^\mu\) with the Noether charge associated with as done in \cite{Nesterenko:1989pz} and here.

A similar though different shift of the mass spectrum was discussed in 
\cite{Ferrara:1992nm} in the context of non-neutral strings in a magnetic field.
Since the effect starts at the order \((qF)^2\) (recall that the field is given in units of the string tension), then experiments to detect a shift in the mass of hadrons in the presence of magnetic fields are are clearly not practical. Instead, one may consider processes involving also closed strings, and use closed strings to probe the open strings. Since our open strings have total charge zero, there must be processes where both open and closed strings are involved. The closed strings propagate in the flat background and do not interact (directly) with the external field, and for them one can unambiguously declare that \(\eta_{\mu\nu}p^\mu p^\nu\) is the mass squared, and they have the usual spectrum of \(M^2 = \frac{2}{\alp}(N+\tilde N-2)\), and with \(N=\tilde N\) for all states. An examination of interactions between closed and neutral open strings in the presence of the EM field might give a preference to one of the definitions of the mass of the open string over the other.

%%%%%%%%%%%%%%%%%%%%%%%%%%%%%%%%
\section{The operator product expansion} \label{sec:OPE}
%%%%%%%%%%%%%%%%%%%%%%%%%%%%%%%%%%%%%%%

We compute the propagator as the singular part of
\be \langle X^\mu(\tau,\sigma) X^\nu(\tau^\prime,\sigma^\prime) \rangle = T[X^\mu(\tau,\sigma) X^\nu(\tau^\prime,\sigma^\prime)] - \nrmO{X^\mu(\tau,\sigma) X^\nu(\tau^\prime,\sigma^\prime)} \ee
where \(T\) denotes time ordering and \(\nrmO{\,\,}\) normal ordering.

We define the normal ordering of operators in the usual way, by putting annihilation operators to the right of creation operators, such that
\be \nrmO{\alpha_m^\mu \alpha_n^\nu} = \alpha_m^\mu \alpha_n^\nu \qquad \nrmO{\alpha_m^\mu \alpha_{-n}^\nu} = \alpha_{-n}^\nu \alpha_m^\mu \ee
for \(n>0\) and any \(m\). As for the zero modes, one defines
\be \nrmO{x^\mu p^\nu} = \nrmO{p^\nu x^\mu} = x^\mu p^\nu \ee
This is because \(p^\mu\) is also an annihilation operator in the sense that \(p^\mu\vac = 0\), which ensures the vacuum is translation invariant. We also have to define \(\nrmO{x^\mu x^\nu}\), since they do not commute. There an ambiguity in how we define it, although the choice only affects the constant part of the two point function. To make a choice, we use the symmetric option of
\be \nrmO{x^\mu x^\nu} = \frac12(x^\mu x^\nu + x^\nu x^\mu) = x^\mu x^\nu - \frac12 [x^\mu,x^\nu] \ee
Inserting the mode expansion for \(X\) and using the above definitions as well as the commutators from the previous section, we find that, in terms of \(z \equiv e^{i\frac\pi\ell(\tau-\sigma)}\) and \(\bar z \equiv e^{i\frac\pi\ell(\tau+\sigma)}\)\footnote{Formally we have to perform the Wick rotation of \(\tau\) before defining the complex coordinates.},
\begin{align} X^\mu(z,\bar z) &X^\nu(w,\bar w)-\nrmO{X^\mu(z,\bar z)X^\nu(w,\bar w)} =  \\ \nonumber
=&[x^\mu,x^\nu]-\frac\alp2 \left(\eta^{\mu\nu}\log|z-w|^2 + \left(\frac{1-qF}{1+qF}\right)^{\mu\nu}\log(z-\bar w) + \left(\frac{1+qF}{1-qF}\right)^{\mu\nu}\log(\bar z-w)\right) \end{align}
This is equivalent to the form presented in \cite{Seiberg:1999vs}, which in our notation is
\begin{align} X^\mu(z,\bar z) &X^\nu(w,\bar w)-\nrmO{X^\mu(z,\bar z)X^\nu(w,\bar w)} = i \frac qT F^\mu{}_\rho g^{\rho\nu} - \\ \nonumber
-&\alp\left(\eta^{\mu\nu}\log|z-w| -\eta^{\mu\nu}\log|z-\bar w| + g^{\mu\nu}\log|z-\bar w|^2 + q F^\mu{}_\rho g^{\rho\nu}\log\frac{\bar z-w}{z-\bar w}\right) \end{align}
On the boundary, the real axis, we write \(z = y_1\), \(w = y_2\). Approaching the boundary from the upper half plane \(z = y_1+i\epsilon\), the last term introduces a branch cut as
\be \log\frac{\bar z-w}{z-\bar w} \to \lim_{\epsilon\to0^+}\bigg(\log(y_1-y_2-2i\epsilon)-\log(y_1-y_2+2i\epsilon)\bigg) = \begin{cases} 0 & y_1-y_2>0 \\ 2\pi i & y_1-y_2<0 \end{cases}\ee
In total we can write the boundary propagator
\be G^{\mu\nu}(y_1,y_2) = -\alp g^{\mu\nu} \log(y_1-y_2)^2 + \frac12i\theta^{\mu\nu}(\sgn(y_1-y_2)+1) \label{eq:G_chargeless}\ee
where
\be g^{\mu\nu} =  (\frac{1}{1-q^2 F^2})^{\mu\nu} \qquad i\theta^{\mu\nu} = [x^\mu,x^\nu] = i\frac{q}{T} F^\mu{}_\rho g^{\rho\nu} \ee
In this expression we see how \(g^{\mu\nu}\), which we first defined in our mode expansion of the solution to relate \(\alpha_0^\mu\) to the momentum can now be interpreted as an effective metric seen by the open string in the presence of an external field.

%%%%%%%%%%%%%%%%%%%%%%%%%%%%%%%%%%%%%%%%%%
\section{Energy momentum tensor on the boundary and the vertex operator}\label{sec:EMtensor}
The question we want to answer in this section is what form the energy-momentum tensor takes, as a local operator defined on the boundary, and then how a vertex operator of the correct dimension can be constructed. Classically, and in the bulk, the energy-momentum tensor is unaffected by the charges, and we have the holomorphic and anti-holomorphic currents
\be T(z) = -\frac1\alp \eta_{\mu\nu} \nrmO{\pa X^\mu \pa X^\nu(z)} \qquad \bar T(\bar z) = -\frac1\alp \eta_{\mu\nu} \nrmO{\bar\pa X^\mu \bar\pa X^\nu(\bar z)} \ee
On the boundary, \(z = \bar z = y\) we suggest that the form of \(T(y) = \bar T(y)\) should be
\be T(y) = -\frac1{2\alp} (g^{-1})_{\mu\nu}\nrmB{\pa_y X^\mu \pa_y X^\nu(y)} \label{modtmn} \ee
where \(g^{-1}\) is defined as the inverse of the tensor \(g^{\mu\nu}\) that appears in the boundary propagator.\footnote{Since we lower indices with \(\eta_{\mu\nu}\), \((g^{-1})_{\mu\nu}\) is not the same as \(g_{\mu\nu}\).} In other words, \(g\) enters the boundary energy-momentum tensor as would a background metric. In addition, \(T(y)\) is defined with respect to the boundary normal ordering defined using the boundary propagator, namely through
\be \nrmB{X^\mu(y_1)X^\nu(y_2)} = X^\mu(y_1)X^\nu(y_2) + 2\alp g^{\mu\nu} \log|y_1-y_2| \ee
in accordance with eq. \ref{eq:G_chargeless} for the boundary propagator. The extra factor of \(\frac12\) in the definition is necessary even in the usual Neumann open string case, because of the doubling in the boundary normal ordering. Note that in this section we ignore the \(\theta^{\mu\nu}\) part of \ref{eq:G_chargeless} since it does not affect the OPEs of operators that involve derivatives.

The motivation behind the above definition of \(T(y)\) is to yield the correct OPEs with itself and other operators when computed directly on the boundary, that is for a primary operator \(\mathcal O\), the OPE when both \(T\) and \(\mathcal O\) are placed on the boundary should read \cite{Cardy:1984bb,Cardy:2004hm}
\be T(y_1) \mathcal O(y_2) \sim \frac{2h_{y}}{(y_1-y_2)^2}\mathcal O(y_2) + \frac{2}{y_1-y_2}\pa_{y} \mathcal O(y_2) \ee
where \(h_y\) is the dimension of the operator under rescaling of \(y\).

In addition to \(T(y)\) we would like to examine if the open string vertex operator for the tachyon (with momentum \(k_\mu\)) also requires a change to reflect the presence of the effective metric. We write the two operators in a more general form
\be T(y) = -\frac1{2\alp} t_{\mu\nu}\nrmB{\pa_y X^\mu \pa_y X^\nu(y)} \ee
\be V_k(y) = \nrmB {e^{i v_{\mu\nu}k^\mu X^\nu}(y)} \ee
and write the different OPEs while keeping track of the different tensor contractions. Then the boundary propagator with the ``effective metric'' derived from \(g^{\mu\nu}\) 
%\be X^\mu(y_1) X^\nu(y_2) \sim -2\alp g^{\mu\nu}\log|y_1-y_2| \ee
implies the following forms for OPEs involving \(T(y)\), denoting \(y_{12}\equiv y_1-y_2\),
\be T(y_1) X^\mu(y_2) \sim \frac{2}{y_{12}}t_{\alpha\beta}g^{\alpha\mu}\pa_y X^\beta\ee

\be T(y_1)\pa_y X^\mu(y_2) \sim \frac{2}{y_{12}^2}t_{\alpha\beta}g^{\alpha\mu}\pa_y X^\beta + \frac{2}{y_{12}}t_{\alpha\beta}g^{\alpha\mu}\pa_y^2 X^\beta\ee

\be T(y_1)T(y_2) \sim \frac{2 t_{\alpha\beta} t_{\mu\nu} g^{\alpha\mu}g^{\beta\nu}}{y_{12}^4} - \frac{2/\alp}{y_{12}^2} t_{\alpha\beta} t_{\mu\nu} g^{\alpha\mu}\pa_y X^\beta \pa_y X^\nu - \frac{2/\alp}{y_{12}} t_{\alpha\beta} t_{\mu\nu} g^{\alpha\mu}\pa_y^2 X^\beta \pa_y X^\nu \ee

\be T(y_1)V_k(y_2) \sim \frac{2\alp t_{\alpha\beta}g^{\alpha\mu}g^{\beta\nu}v_{\mu\rho}v_{\nu\sigma}k^\rho k^\sigma}{y_{12}^2}V_k- \frac{2t_{\alpha\beta}g^{\alpha\mu}v_{\mu\nu}}{y_{12}}i k^\nu \pa_y X^\beta V_k\ee

%\be V_{k_1}(y_1) V_{k_2}(y_2) \sim |y_1-y_2|^{2\alp g^{\mu\nu}v_{\mu\alpha} v_{\nu\beta} k_1^\alpha k_2^\beta}V_{k_1+k_2} \ee

All the OPEs involving \(T\) imply that we have to take \(t_{\alpha\beta}g^{\beta\mu} = \delta_{\alpha}^\mu\), or \(t_{\mu\nu} = (g^{-1})_{\mu\nu}\). This ensures, for example that \(\pa X^\mu\) is a primary operator on the boundary with dimension one, and also that the \(TT\) OPE is of the same form as in the bulk.

Specifically we want that the coefficient of the \(y_{12}^{-4}\) term will remain \(\eta^{\mu}_{\mu} = D\), which is the central charge of the bosons, so it can be canceled by the ghost contribution of \(-26\) on the boundary as well as in the bulk. The OPE of the ghost part of the energy momentum tensor with itself on the boundary is assumed to be unchanged by the presence of the external field. This should be a reasonable assumption as the derivation of the Faddeev-Popov ghosts for the open string does not depend on the details of the boundary condition for the bosons \(X^\mu\), only on the fact that there is a boundary.

Explicitly, the ghost OPE
\be b(z_1) c(z_2) \sim \frac{1}{z_1-z_2} + \frac{1}{z_1-\bar z_2}\ee
implies the doubling in the boundary OPEs 
\be b(y_1) c(y_2) \sim \frac{2}{y_{12}} \qquad \ee
In the ghost boundary energy-momentum tensor we will include a factor of \(\frac12\) relative to the bulk definition and replace \(\pa\to\pa_y\),
\be T_g(y) = \frac12\left(\nrmB{(\pa_y b)c} - 2\nrmB{\pa_y(bc)}\right)(y) \ee
Now the total \(T = T_X + T_g\) on the boundary, with \(t_{\mu\nu} = (g^{-1})_{\mu\nu}\), has the correct OPE with itself
\be T(y_1) T(y_2) \sim \frac{2(D-26)}{y_{12}^4} + \frac{4}{y_{12}^2} T(y_2) + \frac{2}{y_{12}}\pa_y T(y_2) \ee
as can be verified by explicit calculations using the boundary OPEs of \(X\) and the ghosts. Similarly we have the expected OPEs of \(T\) with \(X\) and \(\pa X\),
\be T(y_1) X^\mu(y_2) \sim \frac{2}{y_{12}}\pa_y X^\mu(y_2)\,, \qquad T(y_1) \pa_y X^\mu \sim \frac{2}{y_{12}^2}\pa_y X^\mu(y_2) + \frac{2}{y_{12}}\pa_y^2 X^\mu(y_2) \ee

With the choice \(t_{\mu\nu} = (g^{-1})_{\mu\nu}\) we can also read the dimension of the tachyon vertex operator \(V_k\) as
\be h_y(V_k) = \alp g^{\mu\nu}v_{\mu\alpha}v_{\nu\beta} k^\alpha k^\beta \ee
As discussed in section \ref{sec:spectrum} the spectrum includes states with \(\alp g_{\mu\nu} k^\mu k^\nu = 1-N\). In particular the ground state has \(\alp g_{\mu\nu}k^\mu k^\nu = 1\). To have the correct dimension for \(V_k\) the only possibility is \(v_{\mu\nu} = \eta_{\mu\nu}\). Then all operators of the form \((\pa X)^J V_k\) will have dimension \(J+1\) as it is for the Neumann string.

To summarize, we define the boundary energy momentum tensor as
\be T(y) = -\frac1{2\alp} (g^{-1})_{\mu\nu}\nrmB{\pa_y X^\mu \pa_y X^\nu(y)} \ee
and the tachyon vertex operator is defined as
\be V_k(y) = \nrmB {e^{i k \cdot X}(y)} \ee
using the normal dot product with the flat background metric. The latter definition also ensures that the vertex operator transforms correctly under the translation symmetry \(X^\mu \to X^\mu + a^\mu\), by picking up a phase \(e^{ik\cdot a}\). The open string vertex operator does not change its form relative to the string without endpoint charges. It is still the exponential operator for a string propagating in a flat background. The difference is that it is now defined w.r.t. the boundary normal ordering, which does include \(g^{\mu\nu}\). This choice of \(V_k\) is also what leads to a consistent scattering amplitude, as we will see in the next section.

%The OPE of \(V_k\) with itself is then
%\be V_{k_1}(y_1) V_{k_2}(y_2) \sim |y_1-y_2|^{2\alp g_{\mu\nu} k_1^\mu k_2^\nu}V_{k_1+k_2} \ee

%\paragraph{Derivation?}
%The textbook approach [DiFrancesco][Cardy] towards operators on the boundary of the upper half plane is that they are defined through the OPE of an operator with its ``image'' in the lower half plane,
%\be \phi(z)\phi(z^*) \sim \sum_i (z-z^*)^{h_i-2h}\phi_B^{(i)}(y) \ee
%where \(y = (z+z^*)/2\) is the coordinate on the real axis. [eq. 11.46 in the book] Recall that the extension of the definition of \(T(z)\) to the lower half plane is by
%\be T(z) = \bar T (z^*) \qquad \mathrm{Im} z< 0 \ee
%In particular this ensures that \(T(y) = \bar T(y)\) on the boundary. The effective metric in \(T(y)\) should emerge somehow in the interplay of \(T\) and its image in the lower half plane.

%%%%%%%%%%%%%%%%%%%%%%%%%%%%%%%%%%%%%%%%%%%%%%%%%%%%%%
\section{The scattering amplitude} \label{sec:scattering}
\begin{figure}[h]
\begin{center}
\includegraphics[width=0.44\textwidth]{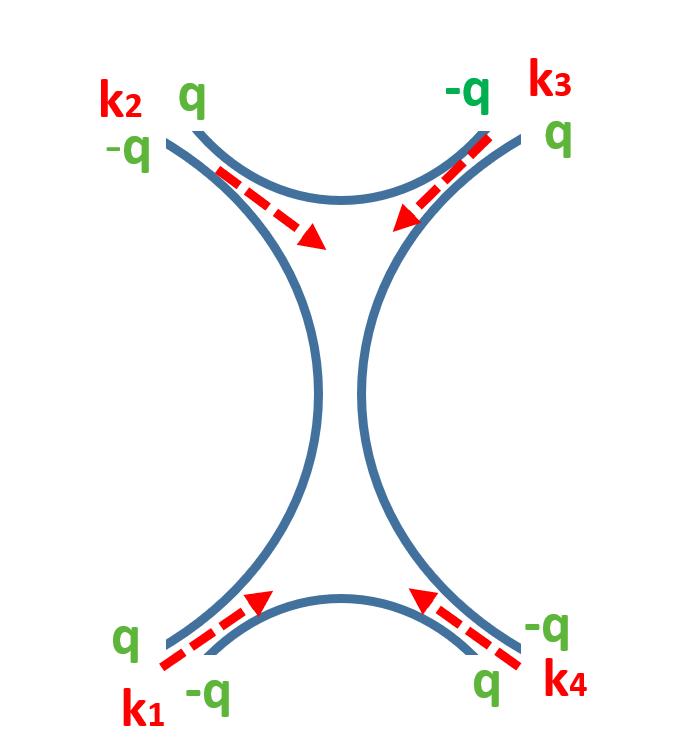}
  \caption{\label{scatchargeless}  The tree level amplitude of a $2\rightarrow 2$ scattering  of strings with $(q, -q)$ charges on their ends. This worldsheet can be mapped to a disk with a charge \(q\) running along its boundary.}
 \end{center}
\end{figure}

Next we consider the scattering amplitude of strings with charges on their endpoints. As in ordinary string scattering  the computation of the level amplitude  translates to calculation of expectations values of vertex operators on a geometry of a disk. The tree level amplitude of strings with the charges $(q,-q)$ on their endpoints is depicted in figure \ref{scatchargeless}. Note that for charge conservation all the strings should have the same charges on their endpoints.

The amplitude is defined as in the ordinary bosonic string theory. We compute it by looking at the expectation value of vertex operators on the disk. The vertex operator was defined in the previous section,
\be V_k(y) = \nrmB {e^{i k \cdot X}(y)} \ee
and it represents an asymptotic tachyon state with incoming momentum \(k^\mu\).

We can use our boundary propagator (we use the result of eq. \ref{eq:G_chargeless}, minus the constant part)
\be G^{\mu\nu}(y_1,y_2) = -2\alp g^{\mu\nu} \log|y_1-y_2| + \frac i2\theta^{\mu\nu}\sgn(y_1-y_2)\ee
to compute the OPE of two exponential operators on the boundary,
\be \nrmO{e^{i k_1\cdot X}(y_1)}\nrmO{e^{i k_2\cdot X}(y_2)} \sim e^{-\frac i2 \theta_{\mu\nu}k_1^\mu k_2^\nu\sgn(y_1-y_2)}|y_1-y_2|^{2\alp k_1 \odot k_2} \nrmO{e^{i(k_1+k_2)\cdot X}(y_2)} \ee
where we have defined the scalar product w.r.t. to \(g_{\mu\nu}\)
\be a \odot b \equiv g_{\mu\nu} a^\mu b^\nu \ee
Since we will write the scattering amplitude for tachyons, all our momenta will obey
\be \alp k_i \odot k_i = 1 \ee
as seen in section \ref{sec:spectrum}. The modifications from the Neumann string case are then a phase from the non-commutativity part, and the modified scalar product in the exponent of \(|y_{12}|\). We can now follow the textbook derivation (see for instance chapter 6 in \cite{Polchinski:1998rq}) of the three and four point functions using our modified propagator.

The expectation value of a product of \(n\) exponents on the boundary of the disk will be
\be \big\langle \prod_{i=1}^n \nrmO{e^{ik_i\cdot X^\mu}(y_i)}\big\rangle_{D_2} = \prod_{i<j} e^{-\frac i2 \theta_{\mu\nu}k_i^\mu k_j^\nu\sgn(y_i-y_j)}|y_i-y_j|^{2\alp k_i \odot k_j} \ee
If momentum is conserved \(\sum_i k_i = 0\), then the phase factor depends only on the cyclic ordering of the \(y_i\), as observed in \cite{Seiberg:1999vs}. For example, for \(n=3\), and using the shorthand notation \(s_{ij} = \sgn(y_i-y_j)\) the phase can be written as
\be -\frac i2 \theta_{\mu\nu}(k_1^\mu k_2^\nu s_{12} - k_1^\mu (k_1+k_2)^\nu s_{13} - k_2^\mu (k_1+k_2)^\nu s_{23}) = -\frac i2 \theta_{\mu\nu} k_1^\mu k_2^\nu (s_{12}+ s_{23}+s_{31} ) \ee
where we have used momentum conservation in the first part, and antisymmetry of \(\theta_{\mu\nu}\) in the second part. The cyclic form of the phase is manifest.

Up to various constants and a delta function for momentum conservation which are the same as in the chargeless case and will be left out of all the following, the amplitude for three tachyons on the disk is
\be S_{D_2}(k_1,k_2,k_3) = \big\langle \nrmO{c^1 e^{i k_1\cdot X}(y_1)} \nrmO{c^1 e^{i k_2\cdot X}(y_2)} \nrmO{c^1 e^{i k_3\cdot X}(y_3)} \big\rangle + (k_2 \leftrightarrow k_3) \ee
The points \(y_i\) have been fixed and integration on each of them has been replaced with the corresponding Faddeev-Popov ghost. We sum over both ways to arrange the three points on the disk by exchanging \(k_2\) and \(k_3\) for the second term.

We assume that the expectation values of the \(c\)-ghosts in the above expression are unaffected by the change of boundary condition and are the same as in the Neumann string case, and we can write the next step as
\be S_{D_2}(k_1,k_2,k_3) = \exp(\frac i2 \theta_{\mu\nu} k_1^\mu k_2^\nu)\prod_{i<j} \vert y_{ij}\vert^{1+2\alp k_i \odot k_j} + (k_2 \leftrightarrow k_3)  \ee
We can use momentum conservation \(k_1+k_2+k_3=0\) and \(\alp k \odot k = 1\) for each of the three states to replace
\be 2 k_1 \odot k_2 = k_3 \odot k_3 - k_1 \odot k_1 - k_2 \odot k_2 = -\frac1\alp \ee
and likewise for other pairs of \(k_i\odot k_j\). The exponent of each \(|y_{ij}|\) vanishes, but we are left with the phases, which combine to give
\be S_{D_2}(k_1,k_2,k_3) = e^{\frac i2 \theta_{\mu\nu} k_1^\mu k_2^\nu} + e^{\frac i2 \theta_{\mu\nu} k_1^\mu k_3^\nu} = \cos(\frac12 \theta_{\mu\nu}k_1^\mu k_2^\nu) \ee
For the last part we again used the antisymmetry of \(\theta_{\mu\nu}\) and momentum conservation to get \(\theta_{\mu\nu}k_1^\mu k_3^\nu = -\theta_{\mu\nu}k_1^\mu (k_1+k_2)^\nu = -\theta_{\mu\nu}k_1^\mu k_2^\nu\). We will use this type of manipulation often in the following. The three point function is independent of the three positions of the points on the disk, which is the first consistency condition. We see that it is a function of the momenta in the plane affected by the external field.

If we proceed to write the four tachyon amplitude on the disk in the same way, again not writing any additional constants, we find that
\begin{align} S_{D_2}(k_1,k_2,k_3,k_4) &= \int_{-\infty}^\infty dy_4 \bigg\langle \prod_{i=1}^3\nrmO{c^1 e^{ik_i\cdot X}(y_i)} \nrmO{e^{i k_4\cdot X}(y_4)} \bigg\rangle + (k_2 \leftrightarrow k_3) \\ 
&= \vert y_{12} y_{13} y_{23} \vert \int_{-\infty}^\infty dy_4 \prod_{i<j}e^{-\frac i2\theta_{\mu\nu}k_i^\mu k_j^\nu \sgn(y_{ij})}\vert y_{ij} \vert^{2\alp k_i\odot k_j} + (k_2 \leftrightarrow k_3) \end{align}
At this point we fix \(y_1 = 0\), \(y_2 = 1\), \(y_3 \to \infty\). We can see again how the mass shell condition \(\alp k \odot k = 1\) removes the dependence on \(y_3\) as we have the factor in the amplitude
\be \vert y_3 \vert^{2+2\alp k_3 \odot (k_1+k_2+k_4)} =\vert y_3 \vert^{2-2\alp k_3 \odot k_3} = 1\ee
We still have to keep track of the phase factors in the integral. As before, the phase depends on the cyclic ordering of the points. We can write the phase explicitly as
\be -\frac i2 \theta_{\mu\nu} \sum_{i<j} k_i^\mu k_j^\nu s_{ij} = -\frac i2 \theta_{\mu\nu}\left[k_1^\mu k_2^\nu(s_{12}+s_{24}+s_{41})+k_1^\mu k_3^\nu(s_{13}+s_{34}+s_{41}) + k_2^\mu k_3^\nu(s_{23}+s_{34}+s_{42})\right] \ee
After choosing \(y_1 = 0\), \(y_2 = 1\), and \(y_3 = \infty\), the phase reduces to
\be i\Theta(y_4) = -\frac i2 \theta_{\mu\nu}\left(k_1^\mu k_2^\nu(-1+\sgn(1-y_4)+\sgn(y_4))+k_1^\mu k_3^\nu \sgn(y_4) + k_2^\mu k_3^\nu \sgn(y_4-1)\right) \label{eq:theta_y}\ee
By integrating \(y_4\) from \(-\infty\) to \(\infty\) and adding the term with \(k_2\) and \(k_3\) exchanged we sum over the six different cyclic orderings of the four vertex operators on the disk,
\be S_{D_2}(k_1,k_2,k_3,k_4) = \int_{-\infty}^\infty dy_4 e^{i\Theta(y_4)}\vert y_4 \vert^{2\alp k_1 \odot k_4} \vert 1-y_4 \vert^{2\alp k_2 \odot k_4} + (k_2 \leftrightarrow k_3) \ee
The integral breaks up into three domains \(y_4 < 0\), \(0 < y_4 < 1\), and \(y_4 > 1\). In each part we will have a different \(\Theta(y_4)\), and each will give a different Beta function multiplied by that phase. The Beta function can be written in one of the two equivalent ways
\be B(a,b) = \int_0^1 dy y^{a-1} (1-y)^{b-1} = \int_0^\infty d\tilde y \frac{\tilde y^{a-1}}{(1+\tilde y)^{a+b}} \ee
We use as before
\be 2\alp k_i \odot k_j = \alp(k_i+k_j)\odot(k_i+k_j) - 2 \ee
and define the modified Mandelstam variables,
\be \tilde s = -(k_1+k_2)\odot(k_1+k_2) \qquad \tilde t = -(k_1+k_3)\odot(k_1+k_3) \qquad \tilde u = -(k_1+k_4)\odot(k_1+k_4) \ee
such that \(\tilde s + \tilde t + \tilde u = -4/\alp\) in the usual way. In terms of \(\tilde s\), \(\tilde t\) and \(\tilde u\) the amplitude is given by
\be\label{VenAmp} S_{D_2}(k_1,k_2,k_3,k_4) = \left[e^{i \Theta_{st}}I(\tilde s,\tilde t) + e^{i\Theta_{su}}I(\tilde s,\tilde u) + e^{i\Theta_{tu}}I(\tilde t,\tilde u)\right] + (k_2 \leftrightarrow k_3) \ee
where the function \(I\) is defined via the Beta function as
\be I(\tilde s,\tilde t) = B(-\alp \tilde s - 1, -\alp \tilde t-1) \ee
The \(st\) part comes from the domain \(y_4>1\), while \(su\) comes from \(y_4<0\) and \(tu\) from \(0<y_4<1\). From the general expression \ref{eq:theta_y} one finds that the phases are
\begin{align}
		\Theta_{st} &= \frac12 \theta_{\mu\nu}(k_1^\mu k_2^\nu - k_3^\mu k_4^\nu) \\
		\Theta_{su} &= -\frac12 \theta_{\mu\nu}(k_1^\mu k_4^\nu - k_2^\mu k_3^\nu) \\
		\Theta_{tu} &= \frac12 \theta_{\mu\nu}(k_1^\mu k_4^\nu + k_2^\mu k_3^\nu) \end{align}
%		An alternative form for the phases is
%		\begin{align}
%		\Theta_{st} &= \frac12\theta_{\mu\nu}(k_1^\mu k_2^\nu - k_2^\mu k_3^\nu + k_3^\mu k_1^\nu) \\
%		\Theta_{su} &= \frac12\theta_{\mu\nu}(k_1^\mu k_2^\nu + k_2^\mu k_3^\nu - k_3^\mu k_1^\nu)  \\
%		\Theta_{tu} &= \frac12\theta_{\mu\nu}(-k_1^\mu k_2^\nu + k_2^\mu k_3^\nu + k_3^\mu k_1^\nu) \end{align}
%The exchange of \(k_2\) and \(k_3\) switches \(\tilde s\) and \(\tilde t\) while the phases transform as  \(\Theta_{st} \to -\Theta_{st}\), \(\Theta_{su}\to -\Theta_{tu}\), and \(\Theta_{tu}\to -\Theta_{su}\).

This leads us to the final answer for the amplitude
\be\label{amplitudefin} S_{D_2}(k_1,k_2,k_3,k_4) = \cos(\Theta_{st}) I(\tilde s,\tilde t) + \cos(\Theta_{su}) I(\tilde s,\tilde u) + \cos(\Theta_{tu}) I(\tilde t,\tilde u)\ee
One can verify that the amplitude is fully crossing symmetric. The amplitude is also invariant under \(q\to-q\), as expected.

\subsection{ Phenomenological implications} \label{sec:pheno}
To relate the scattering amplitude \ref{amplitudefin} to scattering processes of real  neutral   stringy hadrons  like the neutrons or $\pi^0$ one has to derive the analogous result in four dimensions rather than the critical dimensions. This will be discussed in the next section. Meanwhile we would like to examine the amplitude associated with  various different    kinematic setups and  look for ways to confront it with experimental data. 

The reference amplitude is the Veneziano formula with no electric or magnetic background turned on ${\cal A}_0$. Next is the amplitude in the presence of electric field. In that case we distinguish between the case where the direction of the scattering strings is parallel or perpendicular  to the electric field ${\cal A}^E_{||}$ and ${\cal A}^E_{\bot}$ respectively. Similarly, there are two amplitudes when the magnetic field is turned on ${\cal A}^B_{||}$ and ${\cal A}^B_{\bot}$.

 In the case of ${\cal A}_0$ $g^{\mu\nu} = \eta^{\mu\nu}$, $\theta_{\mu\nu}=0$ and the Beta functions are expressed in terms of $s, t$ and $u$.
For example,  ${\cal A}^E_{||}$ for the following kinematic setup\footnote{Note that unlike before, where all the momenta were incoming, now we take \(k_3\) and \(k_4\) to be outgoing.} 
\begin{align}
&k_1 = ( {\cal E},p,0,0)  &\qquad &k_2 = ( {\cal E},-p,0,0) \nonumber \\
&k_3 = ({\cal E},-p\cos\theta,-p\sin\theta, 0)  &\qquad &k_4 = ( {\cal E},p\cos\theta,p\sin\theta,0)
\end{align}
 is given in terms of 
\bea
\tilde s &=& 4( 1-q^2 E^2) {\cal E}^2 \nonumber \\
\tilde t &=& -p^2\left(\frac{(1+\cos\theta)^2}{1-q^2 E^2} +\sin^2\theta\right) \nonumber \\
\tilde u &=& -p^2\left(\frac{(1-\cos\theta)^2}{1-q^2 E^2} +\sin^2\theta\right)
\eea
and 
\be
\Theta_{st}= \frac{qE}{1-q^2 E^2} \frac{{\cal E} p}{T} ( 1+ \cos\theta)   \qquad -\Theta_{su}= \frac{qE}{1-q^2 E^2} \frac{{\cal E} p}{T} ( 1+ \cos\theta)  \qquad \Theta_{tu}=0.
\ee
The other amplitudes  are determined in a similar manner. Note that the effect on the Mandelstam variables starts at quadratic order in the field strength for small values of \(qE\), while the phases are linear in the field. This is true for a general EM field.

The most efficient way to confront the theoretical result for the amplitude of scattering in the presence of constant EM field (\ref{amplitudefin}) is to look for zeros of the amplitude which are not zeros of the amplitude without EM background.
From (\ref{amplitudefin}) it is evident that this can happen if 
\be
\cos(\Theta_{st})  =0,\qquad    \cos(\Theta_{su}) =0 ,\qquad     \cos(\Theta_{tu})  =0
\ee
If instead only $\cos(\Theta_{st})  =0$ holds than the $st$ channel of the amplitude vanishes and similarly for $su$ and $tu$. For the \(st\) amplitude to vanish we need that 
\be
\frac{q}{T} F^\mu{}_\rho \left(\frac{1}{1-q^2F^2}\right)^{\rho\nu} ({k_1}_\mu {k_2}_\nu - {k_3}_\mu {k_4}_\nu) = \pi
\ee 
For the case of a magnetic field $B = F_{12}$  this condition reads 

\be
\frac{q}{T}\frac{B}{1+q^2 B^2} \left[ ({k_1}^1 {k_2}^2)-({k_1}^2 {k_2}^1) - ({k_3}^1 {k_4}^2) +({k_3}^1 {k_4}^2)\right ]= \pi
\ee  
This condition can be obeyed only provided that the expression inside the square brackets does not vanish. If the scattering takes place in a plane perpendicular to either $\hat x_1$ or $\hat x_2$ obviously it does vanish. 
This is also the  case for a center of mass  scattering in the $\hat x_1,\hat x_2$ plane with the momenta 
\be
\vec k_1 = ( k, 0, 0),\qquad \vec k_2 = ( -k, 0, 0),\qquad \vec k_3 = ( \tilde k \cos\phi, \tilde k \sin\phi, 0),\qquad \vec k_4 = ( -\tilde k \cos\phi, -\tilde k \sin\phi, 0) 
\ee
On the other hand for a  scattering in the lab frame of a projectile on a fixed target with
\be
\vec k_1 = ( \tilde k, 0, 0),\qquad \vec k_2 = ( 0, 0, 0),\qquad \vec k_3 = ( k_x, k_y  , 0),\qquad \vec k_4 = ( \tilde k-k_x , -k_y , 0) 
\ee
we find that the condition is 
\be
\frac{q}{T}\tilde k k_y  \frac{B}{1+q^2 B^2}  = \pi
\ee 
and for non-relativistic scattering we have $\tilde k k_y = \tilde k k \sqrt{1- \left ( \frac{k}{\tilde k} \right )}$. This equation determines the value of $k$ for which give $\tilde k$ the channel  $st$ vanishes. In fact it is easy to check that for this kinematic setup this condition leads also to the vanishing of  $\cos(\Theta_{su})$ and $ \cos(\Theta_{tu}) $  and hence of the full scattering amplitude.

Since we write the magnetic field in units of the string tension, then the effect is always very small for realistic values of the magnetic field. The typical string tension for hadrons is \(T \approx 0.18\) GeV\(^{2}\), while the strongest magnetic fields producible in a lab are of the order \(10^1\) Tesla, which is \(\approx 2\times 10^{-15}\) GeV\(^2\). The effect of the magnetic field is also proportional to \(k\tilde k/T\) and can be increased by going to higher and higher transverse momenta, but overcoming a factor of \(10^{-14}\) to reach the point where the amplitude vanishes is unrealistic. 

As an aside, we note that the strongest magnetic fields in nature are in magnetars, a type of neutron star, where fields as large as \(10^{11}\) Tesla \(\sim 10^{-5}\) GeV\(^{2}\) can exist \cite{Duncan:1992hi,Olausen:2013bpa}. Since neutrons are strings with charged particles on their endpoints, one can hope to find some way of making the string model applicable there. Of course this would require a lot of work beyond the scope of this paper. Very strong magnetic fields are also found in heavy ion collisions, in experiments such as RHIC, but there the interaction is with quark-gluon plasma, rather than hadrons \cite{Gursoy:2014aka}.

In real world hadrons there is also the interaction between the charges of the quarks that needs to be taken into account. If we want the effect of the self interaction to be weaker than the external field then as a first order approximation we can write the condition
\be |qF| > \sqrt{\frac{\alpha}{4\pi}} \frac{q^2}{TL^2} \ee
where \(\alpha = \approx \frac{1}{137}\) is the electromagnetic fine structure constant and the charges are given in units of the electron charge, so they are either \(\pm\frac23\) or \(\pm\frac13\) for neutral hadrons. The fields are again given in units of the tension, and we can estimate \(TL^2\) for hadrons as follows. For a rotating string \(J - a = \frac{\pi}{8}TL^2\),\footnote{In our previous works we argue that endpoint masses of the string modify this relation \cite{Sonnenschein:2016pim}, but we neglect them here.} where \(J\) is the angular momentum and \(a\) the intercept of the Regge trajectory. A typical value of the intercept is \(-0.5\), as for the \(\rho\) meson trajectory. Then for the \(\rho_0\) meson for example, the RHS of the last equation can be estimated as (taking \(|q| = \frac23\) even though the \(\rho\) is a combination of \(u\bar u\) and \(d\bar d\)),
\be |F| > \frac{\sqrt{\pi\alpha}}{16}\frac{|q|}{J-a} \approx 0.0126 \ee
We have taken \(J-a \approx 0.5\) for the lowest \(\rho\) meson, but one can relax the condition by going to higher spin mesons, which correspond to longer strings.
With the above condition in mind, even the large magnetic fields discussed above are not as large as the fields one could expect to find in the hadron, so it is not clear if the effects of coupling the string to an external field can be observed for real world hadrons.

\section{The non-critical string with endpoint charges} \label{sec:noncritical}
To relate the scattering amplitude \ref{amplitudefin} or any other of the calculated values in the previous sections to the physics of real world hadrons one has to consider strings in four spacetime dimensions, rather than the critical \(D=26\). 

A framework in which this can be done is that of effective string theory \cite{Polchinski:1991ax,Aharony:2010cx,Aharony:2013ipa,Dubovsky:2012sh,Hellerman:2014cba}. The effective string theory description requires us to expand around a long string, a classical solution of length \(L\). We can then write an effective action for the transverse fluctuations around the long string, where successive terms in the action are of increasing order of the expansion parameter \(\ell_s/L\), with the intrinsic length scale being \(\ell_s = \sqrt{\alp}\). The effective string action gives a consistent description for the string action in any dimension \(D\leq26\).

We use the prescription of Polchinski and Strominger (PS) \cite{Polchinski:1991ax}, where the action is written in the orthogonal gauge. The leading order term is the Nambu-Goto action, and the first added term is
\be S_{PS} = \int d\tau\mathcal L_{PS} = \frac{26-D}{24\pi}\int d\tau d\sigma \frac{(\pa_+^2 X\cdot \pa_- X)(\pa_-^2 X\cdot \pa_+ X)}{(\pa_+ X\cdot \pa_- X)^2} \ee
The coefficient is chosen is such a way that the conformal anomaly is canceled (to the relevant order in the \(1/L\) expansion) for any \(D\). We do not add any new terms to the boundary.

Since the expansion requires us to start from a solution with finite length, we will examine the expansion around the rotating solution that we had for the string in a constant magnetic field (section \ref{sec:rotating}).

\subsection{Spectrum of fluctuations around a rotating solution}
In this section we examine the quantization of the fluctuations around a rotating string in a magnetic field and compare it to the quantization presented in previous sections, where no such expansion was carried out. The analysis of fluctuations around rotating strings without the background magnetic field was addressed in \cite{Frolov:2002av,Baker:2002km,Kruczenski:2004me,Hellerman:2013kba,Zahn:2013yma,Zahn:2016bam}. In a preceding paper we have performed a similar analysis for a rotating string with massive endpoints \cite{Sonnenschein:2018aqf} in a similar manner to what is presented here. Our computation for the string in a magnetic field differs from those in the above references by the addition of the phase \(\phi\) in the solution, implying the presence of a folding point as discussed in section \ref{sec:classical}.

We insert into our action \(X^\mu = X^\mu_{rot} + \delta X^\mu\) where \(X^\mu_{rot}\) is the rotating solution of eq. \ref{eq:rotsol}. Instead of making the orthogonal gauge choice, we gauge away the time component for the fluctuations, meaning \(\delta X^0 = 0\), or \(X^0 = e\tau\). This is the static gauge which fixes \(\tau\), while \(\sigma\) is essentially fixed by the choice of the function \(R(\sigma) = \frac{1}{\omega}\cos(\omega\sigma+\phi)\) used to define the solution.

The full bulk action is the Nambu-Goto action plus PS term as discussed above. To get the leading order terms in the \(1/L\) expansion in the effective action for the fluctuations we take only the terms quadratic in the fluctuations from the NG part of the action. The PS term need only be evaluated on the classical solution in the leading order.

All the fluctuations transverse to the plane of rotation are just free fields. There is no difference for them between the rotating and the non-rotating string, and they are unaffected by the magnetic field. The modes in the plane of rotation are different. 

In polar coordinates \(X^1 = \rho \cos\theta\), \(X^2 = \rho\sin\theta\), we have
\be \rho = \frac{e}{\omega}\cos(\omega\sigma+\phi) + \delta\rho \qquad \theta = \omega\tau + \delta\theta \ee
The mode \(\delta\rho\) is longitudinal to the string, and contributes only boundary terms to the action, while the mode \(\delta\theta\) needs to be redefined to give a canonically normalized kinetic term. This is a mode transverse to the string which we call the planar mode. In terms of
\be f_r = \delta\rho \qquad f_p = \frac{e}{\omega}\cot(\omega\sigma+\phi)\delta\theta \ee
The bulk action for the fluctuations is
\be S = T\int d^2\sigma \left(\frac12 \dot f_p^2 - \frac12f_p^{\prime2}-\frac{\omega^2}{\sin^2(\omega\sigma+\phi)}f_p^2\right) \ee
On the boundary we also have the radial mode. The action is
\be S_b(0) = \frac{T q B}{2}\int d\tau \left(\frac{\omega}{\sin^2\phi}(f_p^2+f_r^2)+\frac{1}{\sin\phi}(f_r \dot f_p-f_p\dot f_r)\right) \ee
\be S_b(\ell) = -\frac{T q B}{2}\int d\tau \left(\frac{\omega}{\sin^2\phi}(f_p^2+f_r^2)-\frac{1}{\sin\phi}(f_r \dot f_p-f_p\dot f_r)\right) \ee

%The variation of \(f_p\) yields the bulk equation of motion
%\be f_p^\dprime - \ddot f_p - \frac{2\omega^2}{\sin^2(\omega\sigma+\phi)}f_p = 0  \ee
%and the boundary equation
%\be f_p^\prime + \frac{q B \omega}{\sin^2\phi} f_p \mp \frac{q B}{\sin\phi}\dot f_r = 0, \qquad \sigma = 0, \ell \ee
From the variation of \(f_r\) we have the boundary condition
\be \frac{q B \omega}{\sin^2\phi} f_r \pm \frac{q B}{\sin\phi}\dot f_p = 0, \qquad \sigma = 0, \ell \ee
which we can immediately solve for \(f_r\).

%Solving the last equation and inserting the solution into the previous one, we find that the boundary condition for \(f_p\) is
%\be f_p^\prime + \frac{q B \omega}{\sin^2\phi} f_p + \frac{q B}{\omega}\ddot f_p = 0, \qquad \sigma = 0, \ell \ee

For \(f_p\) we write the Fourier expansion
\be f_p(\tau,\sigma) = \alpha_0^p f_0(\sigma) + i\sqrt{\frac\alp2}\sum_{n\neq0} \frac{\alpha^p_n}{\omega_n} e^{-i\pil\omega_n\tau}f_n(\sigma) \ee
The bulk equation of motion for the Fourier modes, in terms of the variable \(x = \omega\sigma\), is
%We have to satisfy (for \(x=\omega\sigma\))
\be f_n^\dprime(x) + \left(\omega_n^2-\frac{2}{\sin^2(x+\phi)}\right)f_n(x) = 0 \label{eq:planar}\ee
with the boundary condition
\be f_n^\prime(x) + \tan\phi\left(\frac{1}{\sin^2\phi} - \omega_n^2\right)f_n(x) = 0 \qquad x = 0,\pi \ee
Where we have used the solution for \(f_r\) in order to reach the last equation, as well as \(\tan\phi = qB\). 

The bulk equation can be solved most generally by switching variables to \(y = \cos(x+\phi)\), and inserting \(f_n = (1-y^2)^{1/4}g_n(y)\). Then the equation for \(g_n\) is the Legendre equation, for which the general solution is given by the associated Legendre functions \(P_\nu^\mu\) and \(Q_\nu^\mu\) with \(\mu = 3/2\) and \(\nu = \omega_n-\frac12\). Here it is simpler to write the solution directly in terms of \(x\).

The general solution that satisfies the two boundary conditions has a rather simple form, and it is
\be f_n(x) = N_n \left(\cos(nx)\cot(x+\phi)+n\sin(nx)\right) \ee
The spectrum of eigenfrequencies being
\be \omega_n = n \ee
as before. Note that \(n=0\) also a gives a valid solution, corresponding to a constant shift in \(\theta\).

There is a subtlety to deal with now since the solutions diverge at the folding point \(x+\phi = \pi\). Before we had the magnetic field, this divergence was at the endpoints of the string, and was dealt with by appropriate boundary terms. Without any regularization, the functions \(f_n\) are not normalizable. From standard Sturm-Liouville theory, the eigenfunctions should satisfy the orthonormality relation (for \(m\neq n\))
\be (m^2-n^2)\int_0^\pi dx f_m f_n - \left(f_m f_n^\prime-f_n f_m^\prime\right)\vert_{0}^\pi = 0\label{eq:SturmLiouvilleOrtho} \ee
Using the explicit form of the boundary equation, we can use the above to write
\be \int_0^\pi dx f_m f_n + \tan\phi f_m f_n\vert_{0}^\pi = \pi(\delta_{m+n}+\delta_{m-n}) \label{eq:set}\ee
This equation also defines the measure with respect to which we normalize our eigenfunctions.
Since \(f_n(\pi) = (-1)^n f_n(0)\) the second term on the LHS drops when \(m+n\) is even, and in particular for \(m=n\).

With no magnetic field the eigenmodes diverge at the endpoints, which can be resolved by adding appropriate counter terms on the boundary. The problem of the folding point is that it takes the divergence to some point in the bulk of the string, where no natural counterterms can be written to deal with the divergence.

One possible solution is to introduce a massive particle on the folding point so that it will no longer move at the speed of light. Then the eigenmodes have are finite at the folding point. Then we can renormalize by taking the massless limit properly in a manner similar to that presented in \cite{Sonnenschein:2018aqf} for the string with massive endpoints. This procedure will be described in detail in a follow up paper \cite{Sonnenschein:Folded}.

The starting point is a classical rotating solution for the system that has a mass at the point where the string is folded. If the fold is at the point \(\sigma = s\) and we insert the usual massive particle worldline term into the action, fixed at that point, then in addition to the boundary condition at the endpoints we have the condition at the folding point
\be T (X^{\prime\mu}|_{\sigma=s^-}-X^{\prime\mu}|_{\sigma=s^+}) + m \pa_\tau\left(\frac{\dot X^\mu}{\sqrt{-\dot X^2}}\right) = 0 \ee
We can solve the equations by taking a rotating solution of the same form as before (eq. \ref{eq:rotsol}), with a phase jump at the folding point. That is, we will have
\be \rho(\sigma) = \begin{cases}
\frac{e}{\omega}\cos(\omega\sigma+\phi_1) & \sigma \leq s \\ \frac{e}{\omega}\cos(\omega\sigma+\phi_2) & \sigma \geq s \end{cases} \ee
The phases are related by \(\phi_2 = -\phi_1-2\omega s\) so the function is continuous at the folding point. The two parameters \(\phi_1\) and \(\omega s\) are related by the two boundary conditions and the equation at the folding point.

The physical effect of the mass is to slow down the folding point to a finite velocity as given by the force equation at the fold
\be \frac{2 T}{\gamma_s} = m\omega \gamma_s \beta_s \ee
The massless limit \(m\to 0\) has \(\beta_s = 1\), so \(\gamma_s\to\infty\). From the previous equation \(\gamma_s^2 m \to 2T/\omega\) should be finite at zero mass.

For the fluctuations, we will have to add at \(\sigma=s\) the action
\be S_b(s) = \gamma_s m \int d\tau \left(\frac12 \dot f_p^2 + \frac12\dot f_r^2+\frac12 \mu_p^2 f_p^2+\frac12\mu_r^2 f_r^2+c f_r \dot f_p\right)\ee
with the parameters
\be \mu_p^2 = \gamma_s^2\omega^2\,, \qquad \mu_r^2 = (2\gamma_s^2-1)\omega^2 \,, \qquad c = 2\gamma_s \omega\ee
We will have an additional boundary condition on the fluctuations at the point,
\be f_n^\prime|_{s^+} - f_n^\prime|_{s^-} + \frac{\gamma_s m}{T\omega}(\omega^2\omega_n^2+\mu_p^2 - \frac{c^2\omega^2\omega_n^2}{\omega^2\omega_n^2+\mu_r^2})f_n = 0\,, \qquad x=\omega s\ee
We require that \(f_p\) itself is continuous at the fold, but there is a discontinuity in the derivative \(f_p^\prime\). The allowed eigenfunctions will now be a function of the mass and the magnetic field, that in the massless limit reduces back to \(\omega_n = n\).

With the added boundary condition, we get that we should normalize the eigenfunctions according to
\begin{multline}
\int_0^{\omega\ell} dx f_m f_n + \tan\phi \left(f_m f_n|_{x=0}-f_m f_n|_{x=\omega\ell}\right) +\\+ \frac{\gamma_s m \omega}{T}\left(1-\frac{c^2\omega^2\mu_p^2}{(\omega^2\omega_m^2+\mu_r^2)(\omega^2\omega_n^2+\mu_r^2)}\right)f_m f_n|_{x=\omega s} = \pi (\delta_{m-n} + \delta_{m+n}) 
\end{multline}
In \cite{Sonnenschein:Folded} we show that with the added term at the fold the divergences cancel out and one can take the massless limit.

If one goes ahead and assumes that the subtlety at the folding point has been dealt with, and that by that we have defined a complete set of normalizable eigenfunctions satisfying eq. \ref{eq:set}, then taking the worldsheet Hamiltonian
\be H = T\int_0^\ell d\sigma\left(\frac12\dot f_p^2 + \frac12f_p^{\prime2}+\frac{\omega^2}{\sin^2(\omega\sigma+\phi)}f_p^2\right) + \frac{T\omega}{\sin(2\phi)}\left(f_p^2(0)+f_r^2(0)-f_p^2(\ell)-f_p^2(\ell)\right) \ee
and inserting the mode expansion will give the result
\be \frac{\ell}{\pi} H = \sum_{n=1}^\infty\left(\alpha_{-n}^p\alpha_n^p + \alpha_{-n}^i \alpha_n^i\right) \label{eq:Hrot}\ee
where \(i = 3,\ldots,D-1\) are the transverse directions. Since \(\omega_n = n\) for all modes we find the same spectrum around the rotating solution as we did before when we performed the full quantization of the theory. The same result can be obtained by solving the system with a finite mass at the folding point, and taking the massless limit only at the end.

The Hamiltonian for the fluctuations around the rotating string is then essentially the same as the one found in section \ref{sec:spectrum}. However, we do not see the structure of the zero modes and the non-commutativity in this analysis.

Since we are looking at the spectrum of transverse excitations around the string, we do not have any momentum, at least not in the plane of rotation. Without momentum we cannot hope to see the effective metric in this case. As for the non-commutativity of \([x^1,x^2]\neq0\), for our classical solution we can certainly choose to place our string at the origin \(x^1=x^2=0\). In our solutions for the modes in the plane of rotation \(f_p\) and \(f_r\), we have only the zero mode \(\alpha_0^p\), which amounts to a constant shift in the angle \(\theta\) that does not move the center of mass. One way to see if the semi-classical expansion of the rotating string retains in some form the non-commutativity is to compute the two point function \(\langle X^1 X^2 \rangle\) to see if there is a phase as in the propagator without rotation (eq. \ref{eq:G_chargeless}), which is a more physical manifestation of the non-commutativity than the commutator of the zero modes. We leave that as an open question, and turn to the computation of the intercept in the non-critical rotating string.

\subsection{The intercept} \label{sec:noncrita}
Next we would like to determine the intercept associated with the folded rotating string solution in non-critical dimensions, in the presence of the magnetic field. A more general and complete analysis of rotating strings with folds, which includes both the open string in electromagnetic background and folded closed strings will appear in \cite{Sonnenschein:Folded}. Here we summarize the results for the former case.

The intercept in the semiclassical expansion around the rotating string solution can be defined as
\be a = \langle J - J_{cl}(E) \rangle \ee
where \(J\) is the angular momentum and \(J_{cl}(E)\) the function relating \(J\) and the energy of the string at the classical level. For the ordinary open string \(J_{cl}(E) = \alp E^2\). In terms of the fluctuations, one can show \cite{Sonnenschein:2018aqf} that
\be a = \langle \delta J - \frac{1}{\omega}\delta E \rangle = -\frac1\omega\langle H_{ws} \rangle \ee
where \(H_{ws}\) is the worldsheet Hamiltonian of the fluctuations and \(\omega\) the angular velocity of the string.

The intercept has a contribution of \(\frac{1}{24}\) from each of the \(D-2\) modes transverse to the string, including the planar mode discussed in the previous subsection.

For the non-critical string, in the leading order we have in addition to the result of eq. \ref{eq:Hrot} a constant term in the Hamiltonian that we get by inserting the classical solution into the Polchinski-Strominger term in the action,
\be E_{PS} = -\int_0^\ell d\sigma \mathcal L_{PS}(X_{rot}) = \frac{26-D}{24\pi}\omega\int_0^\pi dx \cot^2(x+\phi) \ee
%For the string rotating in a magnetic field, we have a contribution to the intercept from the PS term that is
%\be a_{PS} = -\frac1\omega\int \mathcal L_{PS}(X_{cl}) = \frac{D-26}{24\pi}\int_0^\pi dx\cot^2(x+\phi)\ee
%\be a_{PS} = -\frac1\omega\int \mathcal L_{PS}(X_{cl}) = \frac{D-26}{24}\frac1\ell\int_0^\ell d\sigma \cot^2(\pil\sigma+\phi) = \frac{D-26}{24\pi}\int_0^\pi dx\cot^2(x+\phi)\ee
This term will contribute a correction to the intercept for \(D\neq26\). The integral diverges, with the source of the divergence being the folding point \(x+\phi=\pi\) that moves at the speed of light. For the string with no charges the divergence can be regularized by appropriate boundary terms as counterterms, which are a geodesic curvature term or massive endpoints of the string \cite{Baker:2002km,Hellerman:2013kba,Zahn:2016bam,Sonnenschein:2018aqf}. For the rotating solution in a magnetic field, it is the folding point, rather than the endpoints of the string, that moves at the speed of light, and consequently we cannot regularize the divergence by adding boundary terms.

We again turn to the solution with a massive particle on the folding point described in the previous subsection. Evaluating the PS Lagrangian on this solution, we find the same form as for the string with massive endpoints,
\be E_{PS} = \frac{26-D}{12\pi}\left(\frac{4 T}{\gamma m}-\frac{4(\arcsin\beta)^2}{\tilde L}\right)\ee
where here the relevant velocity \(\beta\) is that of the massive particle at the folding point.
As explained in \cite{Sonnenschein:2018aqf}, we can write the energy in terms of the effective length of the rotating string,
\be \tilde L = L \frac{\arcsin\beta}{\beta} \ee
and the relativistic mass \(\tilde m = \gamma m\), since that is the form that appears in the classical energy of the rotating string solution
\be E = \gamma m + T L \frac{\arcsin\beta}{\beta} = \tilde m + T\tilde L  \ee
For the string with massive endpoints it was argued that we can redefine the mass at the endpoint of the rotating string to eliminate the first term in \(E_{PS}\) that diverges in the massless limit, and that we get the correct finite result for the intercept by considering only the \(1/\tilde L\) term. If we do this here, we find that the subtraction can be done without reference to the value of the magnetic field, so that
\be a_{PS} = -\frac1\omega E_{PS}^{(ren)} = \frac{26-D}{24} \ee
as in \cite{Hellerman:2013kba,Sonnenschein:2018aqf} in the massless limit even when the string is in a magnetic field. This means that the total intercept, considering all transverse fluctuations as well the PS contribution remains
\be a = \frac{D-2}{24} + \frac{26-D}{24} = 1 \ee
for any \(D\) and any value of the magnetic field \(B\).

\subsection{Comments on the scattering amplitude for non-critical strings}
In the previous section we discussed the spectrum and the intercept at non-critical dimensions. 
We would like next to address the same issue about the scattering amplitude, namely, the vertex operators and the expectation values of their products. 

In \cite{Hellerman:2017upi} off-shell vertex operators in effective string theory in non-critical dimension were written down. It is emphasized there that these vertex operators  correspond to external background fields rather than states of the string theory itself. For our purposes this is not what is needed. Instead we need the on-shell vertex operator for the scalar state of the open string.\footnote{In the ordinary string theory this is obviously a tachyonic state but in the string theory of hadrons it is not since there are such hadronic states like $\pi^0$. This can achieved if in that theory the intercept is negative \cite{Sonnenschein:2017ylo}.} Nevertheless, we would like to examine the proposal of
\cite{Hellerman:2017upi}. The boundary vertex operator of a scalar state in non-critical dimensions is expected to take the form 
\be
 V_k = \nrmO{e^{i \eta_{\mu\nu}k^\mu X^\nu} e^{\gamma \varphi}(y)} 
\ee 
where $\varphi$ is the Liouville field and $\beta$ should be determined from the requirement that the operator have dimension one. It was shown in \cite{Polchinski:1991ax,Hellerman:2014cba} that the Liouville field can be expressed as a composite operator of the following form
\be
\varphi = -\frac12 \log\left ( \pa_\alpha X^\mu \pa^\alpha X_\mu\right ) 
\ee
Following the discussion of section (\ref{sec:EMtensor}) the space time metric 
in this expression should be $\eta_{\mu\nu}$.  Thus, the vertex operator takes the form
\be V_k = \nrmO{e^{i \eta_{\mu\nu}k^\mu X^\nu} \left( \pa_\alpha X^\mu \pa^\alpha X_\mu\right )^{-\gamma/2} (y)}  \ee

However, the world sheet energy momentum tensor that determines $\beta$ should have the modified metric as in 
(\ref{modtmn}). In addition there is a contribution of the Liouville term  so the boundary energy momentum tensor reads
\be T(y) = -\frac1\alp (g^{-1})_{\mu\nu}\nrmB{\pa_y X^\mu \pa_y X^\nu(y)} +\frac{D-26}{24}\left(\frac{(g^{-1})_{\mu\nu}\nrmB{\pa_y X^\mu \pa^3_y X^\nu(y)}}{(g^{-1})_{\mu\nu}\nrmB{\pa_y X^\mu \pa_y X^\nu(y)}} \right )\label{modtmnnoncritical} \ee
 In \cite{Hellerman:2017upi} it was found out that to determine $\gamma$ to leading order one has to use only the first term in the energy momentum tensor and at that order the vertex operator reads 
\be V_k = \nrmO{e^{i \eta_{\mu\nu}k^\mu X^\nu} \left( \pa_\alpha X^\mu \pa^\alpha X_\mu\right )^{-\alp k \odot k +1} (y)}  \ee
The conclusion is thus that to leading order since on shell $\alp k \odot k =1$ the Liouville dressing is dropped off and the vertex operator is the one of the critical dimensions. In fact also the first order correction  to the vertex operator found in \cite{Hellerman:2017upi} vanishes on shell. Hence it seems that to leading order in the string length the scattering amplitude of the non-critical string is the same as that of the critical string.

%%%%%%%%%%%%%%%%%%%%%%%%%%%%%%%%%%%%%%%%%%%%
\section {Summary} \label{sec:summary}
In this paper we have determined the scattering amplitude of an open string with two opposite charges on its endpoints in a constant electromagnetic background field. As part of the introduction we have already summarized the results of this work. These include: (i) Classical solutions of rotating strings requiring folds.  (ii) The modification of the spacetime metric along the directions of the EM field. (iii) The non-commutativity of the center of mass coordinates along those directions. (iv) The determination of the scattering amplitude for this type of strings, including the Mandelstam variables that are changed due to the modified metric and certain phases that follow from the non-commutativity parameters. (v) The fact that the quantum  energy momentum on the boundary of the string worldsheet is affected by the boundary charges and is different from the Noether procedure result. (vi) Phenomenological implications including the observation that by tuning the magnetic field or the kinematic parameters the scattering amplitude can be set to zero. (vii) Mapping the scattering amplitude of strings in the critical dimensions to the case of non-critical four dimensions. 

There are several open questions that follow from the results of the current paper. Among them are:
\begin{itemize}
\item
The second part of this project is the determination of the scattering amplitude of a string with charges at the endpoints of an open string for the general case where the total charge of the string is non-zero, \(q_1+q_2 = Q\neq0\). This generalization is highly non-trivial, in part due to the fact that the general charges now implies different boundary conditions for the two endpoints of the string. In section \ref{sec:modeexp} we showed how that implies a constant shift in the spectrum. The charged string case raises several other questions, for instance if there is a quantization condition on the endpoint charges, and the question of how one expands around an accelerating string if there is an electric field. All these questions will be addressed in a sequel publication \cite{Sonnenschein:SACSgeneral}.

\item
As we have already mentioned part of the motivation for this analysis is to determine physical experiments that will verify the stringy nature of hadrons. Here in this note we took a constant electromagnetic field to couple to the open strings. In a similar manner one see how the string is affected by non-constant fields, like for instance electromagnetic plane waves or other types of radiation fields.

\item
Hadronic strings in the form of open strings with massive and charged endpoints differ from the ordinary string not only when coupled to an external electromagnetic fields but also due to the mutual electromagnetic interaction between the two endpoints. An analysis of the corresponding quantum spectra of such strings and scattering amplitudes will be presented in future work.

\item
Part of the motivation for the present paper has been to develop new ways to be sensitive to the stringy nature of hadrons. In addition to the spectrum we would like to have more measurable quantities that will  enable us  to reveal the stringy  nature of hadrons. Form factors like magnetic moments and structure functions  are  different  for hadrons as composite particles and hadrons as strings.

\item
The generalization of Veneziano's amplitude to the scattering amplitude of strings in four or any other non-critical dimensions has not been investigated thoroughly. Certain issues have been worked out in  \cite{Hellerman:2017upi} when considering the ordinary open string. We have made several basic comments about  the case of strings with charges on their endpoints. Further exploration of this issue  in both the ordinary and charged strings is an obvious open question.

\item
A very important challenge of any stringy determination of scattering amplitudes of hadrons is to be able to incorporate the ``bending"  of the Regge trajectories \(\alpha(t)\) at very large negative $t$, which follows from the asymptotic freedom of QCD even at large $N_c$.\footnote{We thank the referee for pointing this issue to us.} This issue will be the subject of our future research work.

\item
For the hadronic strings which have massive endpoints, or the string in the magnetic field, it is natural to expand around a rotating solution. We have shown that quantizing the fluctuations around the rotating strings gives the same spectrum that we get without the expansions. However, the structure of zero modes is different and this needs further exploration. Another important question is seeing how such a semi-classical expansion can be used to compute scattering amplitudes for long non-critical strings.

\item Another interesting feature of the rotating strings is that they can develop folds, and the rotating solution in a magnetic field requires at least one fold. These folds introduce certain divergences, and we have given a short explanation here of how they can be regularized by introducing a mass at the fold. This treatment can be generalized to other solutions with folds, specifically to rotating closed strings.

%\item
%A challenge that goes with us all along the odyssey in the land of stringy hadrons is obviously the derivation of the string effective action from the underlying QCD theory of quarks and gluons. 
\end{itemize}
\appendix
%\bibliographystyle{JHEP}
%\bibliography{Biblio}
%\section{Notation and conventions}

%%%%%%%%%%%%%%%%%%
\section*{Acknowledgments}
We would like to thank Michael Green, Massimo Porrati,  Amit Sever,  Stefan Theisen, Arkady Tseytlin, and Gabriele Veneziano  for useful discussions. We would also like to thank Karthik Inbasekar who took part in the first stages of this project. This work was supported in part by a center of excellence supported by the Israel Science Foundation (grant number 2289/18).  

\bibliographystyle{JHEP}
\bibliography{SACS}

\appendix

\section{Examples and particular cases}
\subsection{Lorentz symmetry} \label{app:Lorentz}
In four dimensions, there are two independent solutions of eq.  \ref{eq:deltaF}, that is two symmetries that remain unbroken after fixing the external field \(F_{\mu\nu}\). In four dimensions, we can always go to a frame where the electric field to is parallel to \(X^1\), then rotating in the \(23\) plane to eliminate the \(X^2\) component of the magnetic field. Then
\be F_{\mu\nu} = \bmat 0 & -e & 0 & 0 \\ e & 0 & b\cos\alpha & 0 \\ 0 & -b\cos\alpha & 0 & b\sin\alpha \\ 0 & 0 & -b\sin\alpha & 0 \emat \label{eq320} \ee
There are always two independent transformations we can make which leave the above \(F_{\mu\nu}\) invariant, which are combinations of rotations and boosts in the most general case. Their generators are given by
\be \omega^{(1)}_{\mu\nu} =
\bmat 0 & 0 & 0 & -b e\cos\alpha \\
0 & 0 & b^2\cos\alpha\sin\alpha & 0 \\ 
0 & -b^2\cos\alpha\sin\alpha & 0 & e^2+b^2\sin^2\alpha \\
be\cos\alpha& 0 & -e^2-b^2\sin^2\alpha & 0 \emat \ee
%
%\be \omega^{(2)}_{\mu\nu} = \bmat
%0 & 1+\sin^2\alpha & -\cos\alpha\sin\alpha\sin\beta & \cos\alpha\sin\alpha\cos\beta \\
%-1-\sin^2\alpha & 0 & \cos\alpha\cos\beta & \cos\alpha\sin\beta \\
%\cos\alpha\sin\alpha\sin\beta & -\cos\alpha\cos\beta & 0 & 0 \\
%-\cos\alpha\sin\alpha\cos\beta & -\cos\alpha\sin\beta & 0 & 0 
%\emat \epsilon \ee
\be \omega^{(2)}_{\mu\nu} = \bmat
0 & e^2+b^2\sin^2\alpha & 0 & b^2\cos\alpha\sin\alpha \\
-e^2-b^2\sin^2\alpha & 0 & be\cos\alpha & 0 \\
0 & -be\cos\alpha & 0 & 0 \\
-b^2\cos\alpha\sin\alpha & 0 & 0 & 0 \emat \ee
In dimensions higher than four the counting of the symmetries is more involved, as the numbers of components of both \(F_{\mu\nu}\) and general Lorentz transformations increase.

\subsection{Particular solutions of the equations of motion} \label{app:Solutions}
Here we write the explicit forms the solutions of \(M_{\mu\nu}\) (eq. \ref{eq:MF}) and the effective metric (\ref{matrixg}) take for some specific cases. We use the metric signature \((-,+,+,+,\ldots)\). In this sign convention the electromagnetic tensor \(F_{\mu\nu}\) in four dimensions is
\be F_{\mu\nu} = \bmat 0 & -E_1 & -E_2 & -E_3 \\ E_1 & 0 & B_3 & -B_2 \\ E_2 & -B_3 & 0 & B_1 \\ E_3 & B_2 & -B_1 & 0 \emat \ee

For a pure magnetic field, \(F_{12} = - F_{12} = B\) and other components zero, the matrix \(M\) relating the left and right moving modes of the open string is in simply a rotation matrix in the 12 plane affected by the magnetic field
\be M = \frac{1}{1+q^2 B^2}\bmat 1-q^2 B^2 & -2q B \\ 2q B & 1-q^2 B^2 \emat = \bmat \cos\alpha & \sin\alpha \\ -\sin\alpha & \cos\alpha \emat \ee
with the rotation angle
\be \sin\alpha = -\frac{2 q B}{1+q^2 B^2} \ee
The effective metric is
\be g^{\mu\nu} = \bmat -1 & 0 & 0 & 0 \\ 
												0 & \frac{1}{1+q^2 B^2} & 0 & 0 \\
												0 & 0 & \frac{1}{1+q^2 B^2} & 0 \\
												0 & 0 & 0 & 1 \emat \ee

For an electric field we have a pure boost in the direction of the electric field
\be M = \frac{1}{1-q^2 E^2}\bmat 1+q^2E^2 & 2q E \\ 2q E & 1+q^2 E^2 \emat = \bmat \gamma & -\gamma\beta \\ -\gamma\beta & \gamma \emat \ee
with the boost parameter
\be \beta = \frac{-2qE}{1+q^2 E^2} \qquad \gamma = \frac{1}{1-\beta^2} = \frac{1+q^2 E^2}{1-q^2 E^2} \ee
and with the effective metric (for \(E = -F_{01}\))
\be g^{\mu\nu} = \bmat -\frac{1}{1-q^2 E^2} & 0 & 0 & 0 \\ 
												0 & \frac{1}{1-q^2 E^2} & 0 & 0 \\
												0 & 0 & 1 & 0 \\
												0 & 0 & 0 & 1 \emat \ee
Here we see clearly the a critical value of the field \(|qE| = 1\), where the metric is singular \cite{Bachas:1992bh}.

Another interesting particular case is when we combine an electric with a transverse magnetic field.
We take \(F_{01} = -E\) and \(F_{12} = B = \alpha E\).
%\be F = E \bmat 0 & 1 & 0  & 0\\ 1 & 0 & -\alpha  & 0\\ 0 & \alpha & 0  & 0 \\  0 & 0 & 0 & 0\emat \ee
If \(|\alpha| < 1\) we can boost the system to a frame with only an electric field, while for \(|\alpha| > 1\) we have a frame with only a magnetic field, so the return to one of the two cases described above. In the special case of \(\alpha = 1\), \(B = E\), the matrix \(M\) is
\be M^\mu{}_\nu = \bmat 1+ 2q^2 E^2 & 2qE & - 2q^2 E^2 & 0 \\ 
												2qE & 1 & -2qE & 0 \\
												2q^2 E^2 & 2qE & 1- 2q^2 E^2 & 0 \\
												0 & 0 & 0 & 1 \emat \ee
while the effective metric takes the special form
\be g^{\mu\nu} = \bmat -1-q^2 E^2 & 0 & q^2 E^2 & 0 \\ 
												0 & 1 & 0 & 0 \\
												q^2 E^2 & 0 & 1-q^2 E^2 & 0 \\
												0 & 0 & 0 & 1 \emat \ee
Here there is no apparent singularity when we reach the critical field \(|qE|=1\).

\end{document}